\documentclass[12pt]{article}
\newcommand{\be}{\begin{equation}}
\newcommand{\ee}{\end{equation}}

\newcommand{\bi}{\begin{itemize}}
\newcommand{\ei}{\end{itemize}}
\newcommand{\bea}{\begin{eqnarray}}
\newcommand{\eea}{\end{eqnarray}}
\newcommand{\ba}{\begin{array}}
\newcommand{\ea}{\end{array}}

\usepackage[left=2.50cm, right=2.50cm, top=2.50cm, bottom=2.50cm]{geometry}
\usepackage[utf8]{inputenc}
\usepackage{cancel}
\usepackage[english]{babel}
\usepackage{physics}
\usepackage{hyperref}
\usepackage{amsthm}
\usepackage{mathrsfs}
\usepackage{mathtools}
\usepackage{graphicx}
\usepackage{changepage}
\usepackage{amssymb,amsmath}
\usepackage{verbatim}
\usepackage{empheq}
\usepackage{xcolor}
\usepackage{tikz}
\usepackage{float}
\usepackage{multirow}
\usepackage{multicol}
\usepackage{amsmath}
\usepackage{centernot}

\usepackage{tensor}
\newcommand{\mixt}[3]{\tensor{#1}{^{#2}_{#3}}}
\newcommand{\invmixt}[3]{\tensor{#1}{_{#2}^{#3}}}

\usepackage[hang,flushmargin]{footmisc} 
\usepackage[normalem]{ulem}
\usetikzlibrary{tikzmark}
\numberwithin{equation}{section}
\hypersetup{hidelinks}

\newlength{\bibitemsep}
\newlength{\bibparskip}
\let\oldthebibliography\thebibliography
\renewcommand\thebibliography[1]{%
  \oldthebibliography{#1}%
  \setlength{\parskip}{\bibitemsep}%
  \setlength{\itemsep}{\bibparskip}%
}
\begin{document}

\par
\bigskip
\Large
\noindent
{\bf 
Covariant field theory of 3D massive fractons
}

\bigskip
\par
\rm
\normalsize

\hrule

\vspace{0.7cm}
\large
\noindent
{\bf Erica Bertolini$^{1,a}$},
{\bf Matteo Carrega$^{2,b}$},\\
{\bf Nicola Maggiore$^{3,4,c}$},
{\bf Daniel Sacco Shaikh$^{3,4,d}$}\\

\par

\small

\noindent$^1$ School of Theoretical Physics, Dublin Institute for Advanced Studies, 10 Burlington Road, Dublin 4, Ireland.

\noindent$^2$ CNR-SPIN, Via Dodecaneso 33, 16146 Genova, Italy.

\noindent$^3$ Dipartimento di Fisica, Universit\`a di Genova, Via Dodecaneso 33, I-16146 Genova, Italy.

\noindent$^4$ Istituto Nazionale di Fisica Nucleare - Sezione di Genova, Via Dodecaneso 33, I-16146 Genova, Italy.
\smallskip

\smallskip

\vspace{1cm}

\noindent
{\tt Abstract}\\

We construct a covariant and gauge-invariant theory describing massive fractons in three spacetime dimensions, based on a symmetric rank-2 tensor field. The model includes a Chern-Simons-like term that plays a dual role: it generates a topological mass for the tensor gauge field and simultaneously acts as a source of intrinsic fractonic matter. This dual mechanism is novel and leads to a propagating fractonic degree of freedom described by a massive Klein-Gordon equation. The theory propagates two degrees of freedom -- one massive, one massless -- whose number is preserved in the massless limit, in analogy with the Maxwell-Chern-Simons mechanism of Deser-Jackiw-Templeton. We analyze the resulting equations of motion and show that the intrinsic fractonic matter satisfies Gauss- and Amp\`ere-like laws, with conserved dipole {and trace of the quadrupole moment}. Upon coupling to external matter, a second fractonic sector emerges, leading to a coexistence of intrinsic and extrinsic subsystems with different mobility and conservation properties. Our model provides a unified framework for describing massive fractons with internal structure, and offers a covariant setting for exploring their interactions and extensions.

\vspace{\fill}

\noindent{\tt Keywords:} \\
Quantum field theory, tensor gauge field theory, fractons, topological mass generation.

\vspace{1cm}

\hrule
\noindent{\tt E-mail:
$^a$ebertolini@stp.dias.ie,
$^b$matteo.carrega@spin.cnr.it,
$^c$nicola.maggiore@ge.infn.it,
$^d$daniel.sacco.shaikh@ge.infn.it.
}
\newpage


\section{Introduction}

Fractons are exotic quasiparticles characterized by unusual mobility constraints: isolated charges are strictly immobile, while certain bound states can move only along restricted subspaces \cite{Nandkishore:2018sel,Pretko:2020cko,Gromov:2022cxa}. These constraints arise intrinsically from generalized conservation laws
 of dipole or higher multipole momenta, rather than from external potentials \cite{Pretko:2016kxt,Pretko:2016lgv,Gromov:2018nbv}. Initially discovered in lattice models in three spatial dimensions featuring subsystem symmetries and constrained Hilbert spaces, such as the Chamon model \cite{Chamon:2004}, the X-cube model \cite{Vijay:2016phm} and the Haah's code \cite{Haah:2011}, fractons have since then inspired a rapidly growing literature due to their rich physical implications and connections to diverse areas, including topological phases of matter, quantum information, elasticity theory, and even quantum gravity \cite{Pretko:2020cko, Ma:2017aog,Brown:2019hxw,Pretko:2017kvd,Gromov:2017vir,Gromov:2019waa,Caddeo:2022ibe,Pena-Benitez:2023aat,Hartong:2024hvs}.
While fractons originated within quantum information theory and condensed matter physics, they have increasingly attracted attention from the high-energy physics community. This is due in part to their unusual symmetry and electromagnetic-like structures involving higher-rank tensor gauge theories and subsystem symmetries \cite{Pretko:2016kxt, Shirley:2018vtc}, but also because they challenge conventional notions of locality, mobility, and gauge invariance in field theory. 
Notably, links between fractonic matter and emergent gravity-like phenomena have been proposed, suggesting intriguing connections to gravitational physics \cite{Pretko:2017fbf}, generalized electromagnetism \cite{Pretko:2016lgv}, and elasticity \cite{Pretko:2017kvd}.
A breakthrough came with Pretko's introduction of non-relativistic effective field theories for fractons based on a symmetric rank-2 tensor gauge field with a generalized Gauss law enforcing dipole conservation \cite{Pretko:2016kxt,Pretko:2016lgv,Pretko:2017xar}. These theories naturally encode fractonic constraints through higher-moment conservation laws and exhibit Maxwell-like properties and excitations with subdimensional dynamics \cite{Pretko:2016lgv}. Subsequent works expanded these ideas by linking fractons to generalized global symmetries \cite{Gaiotto:2014kfa,
 Seiberg:2020bhn,McGreevy:2022oyu,Cordova:2022ruw}  and hydrodynamics \cite{Gromov:2020yoc,  Doshi:2020jso, Grosvenor:2021rrt}, thereby placing fracton physics within an even broader theoretical framework.
More recently, there has been growing interest in developing relativistic and covariant formulations of fracton field theories \cite{Blasi:2022mbl,Bertolini:2022ijb,Bertolini:2023juh, Bertolini:2023sqa,Afxonidis:2023pdq, Bertolini:2024yur,Rovere:2024nwc, Bertolini:2025jul,Bertolini:2025qcy,Bertolini:2024apg,Hinterbichler:2025ost}.  Such formulations enable the adoption of standard quantum field theory tools, and are essential for exploring high-energy applications and couplings to curved backgrounds or external probes.
Most existing models focus on massless fractons, describing gapless phases with long-range correlations;
however, many physical systems exhibit gapped excitations and finite correlation lengths, motivating the study of massive fracton phases. Massive theories can describe softened mobility constraints or bound states and typically allow a clearer identification of the propagating Degrees of Freedom (DoF). In the context of higher-rank lattice gauge theories of gapless fractons, it has been shown that in most cases, by breaking the $U(1)$ gauge symmetry into $\mathbb{Z}_n$  through a Higgs-like mechanism \cite{Bulmash:2018lid, Ma:2018nhd}, the excitations acquire mass, but the fractonic behaviour is lost. The only known exception is the X-cube model \cite{Vijay:2016phm, Slagle:2017wrc}, which is defined in three space dimensions and, being a lattice model, is not covariant and does not rely on a field theoretical set up. On the other hand, non-covariant rank-2 Maxwell-Chern-Simons (MCS) models with fractonic features have been proposed for instance in \cite{Ma:2020svo,Chen:2023oov}, where the massive term has an infinite-dimensional matrix coefficient. Another example appears in \cite{Pretko:2017xar, Prem:2017kxc}, where a Witten-like effect for fracton is considered. In that case, a non-covariant higher-rank theory is briefly mentioned, in which a dynamical traceless term is introduced, in three spacetime dimensions (3D), as a regulator of a Chern-Simons (CS)-like term. In this way an energy gap arises, and massive fractonic excitations emerge in the so-called ``traceless scalar charge theory'' \cite{Pretko:2016kxt,Pretko:2016lgv}. Finally, the possibility of having a mass in fracton models can be relevant to interpret gravitational-like effects \cite{Pretko:2020cko}. In fact, an attractive force can be attributed to massive fracton quasiparticles, derived from an effective potential arising from locality and the conservation of the center of mass \cite{Pretko:2017fbf}. Introducing a mass term in a gauge theory generally breaks gauge invariance unless accompanied by additional fields or topological mechanisms. In 3D, the MCS  theory provides an elegant example: a parity-odd CS term gives mass to a vector gauge field while preserving gauge invariance and unitarity  \cite{Deser:1981wh,Deser:1982vy}. In the context of covariant fracton theories, both Maxwell and CS -like terms has been shown to exist \cite{Bertolini:2022ijb,Bertolini:2024yur} as higher-rank fractonic generalizations of the ordinary ones. This raises a natural question: can a similar topological mass generation mechanism be implemented for higher-rank tensor gauge theories describing fractons~?
Addressing this requires the construction of a covariant gauge theory for a symmetric rank-2 tensor field with a suitable kinetic term and a gauge-invariant topological mass term analogous to the CS action \cite{Bertolini:2024yur}. 
The resulting theory must preserve the gauge symmetry characteristic of fracton models, admit a smooth and physically meaningful massless limit, avoiding discontinuities analogous to the van Dam-Veltman-Zakharov (vDVZ) discontinuity in massive gravity  \cite{vanDam:1970vg, Zakharov:1970cc}, and support well-defined propagators. This issue has been successfully addressed and solved in  \cite{Blasi:2017pkk,Blasi:2015lrg,Gambuti:2021meo,Bertolini:2023wie} for 4D massive Linearized Gravity, which is  the paradigmatic theory of a  symmetric rank-2 tensor field.
Beyond the technical construction, we will show that topological mass terms in fracton theories play a dual role: they not only generate mass but also potentially encode intrinsic fractonic matter and thus give rise to nontrivial fractonic configurations, even without external ordinary matter couplings.
 This reflects a deeper unity between the geometric and macroscopic aspects of the theory, bridging topological field theory and fracton phenomenology.\\
 
 Inspired by the above results, within the context of a covariant field theoretical approach, the present study aims to:

\begin{itemize}
\item
formulate a relativistic and gauge-invariant field theory for massive fractons in 3D;

\item
explore mass generation via a higher-rank CS-like term, in analogy with the MCS mechanism;

\item
preserve the defining features of fracton dynamics, including gauge constraints and generalized conservation laws;

\item
ensure a consistent and continuous massless limit free from pathologies;

\item
provide a solid field-theoretic framework for further studies involving couplings to external matter, gravitational backgrounds, and topological phases.
\end{itemize}

The paper is organized as follows. In Section~2, we introduce the covariant gauge theory for a symmetric rank-2 tensor field in 3D and define the CS-like term. Section~3 sets up the perturbative framework and discusses the structure of the kinetic operator and propagator. In Section~4, we examine the DoF, analyze gauge-fixing conditions, and verify the consistency of the massless limit. Section~5 reformulates the equations of motion (EoM) in terms of generalized electric and magnetic fields and identifies the fractonic constraints they obey. In Section~6, we study the energy-moment tensor and the structure of conserved currents. Section~7 considers the coupling to external matter sources and the resulting modifications of the field equations. We conclude in Section~8 with a summary and possible directions for further investigation.

\vspace{1cm}

{\bf Notations}

3D=2+1 spacetime dimensions .\\
Indexes:  $\mu,\nu,\rho,...=\{0,1,2\}\ \ \ i,j,k,...=\{1,2\}$ .\\
Minkowski metric: $ \eta_{\mu\nu}=\mbox{diag}(-1,1,1)\ . $ \\
Levi-Civita symbol: $\epsilon_{012}=1=-\epsilon^{012}\ $.

\section{The general model}\label{quasitopmassgen}

\subsection{The action}

We consider a 3D symmetric rank-2 tensor field $h_{\mu\nu}(x)$ transforming as \cite{Blasi:2022mbl,Bertolini:2022ijb,Bertolini:2023juh,Bertolini:2023sqa,Bertolini:2024yur,Bertolini:2025jov}
	\be\label{longsimm}
	\delta h_{\mu\nu}=\partial_\mu\partial_\nu\phi\ ,
	\ee
which is the infinitesimal diffeomorphism transformation
	\be
	\delta h_{\mu\nu}=\partial_\mu\phi_\nu + \partial_\nu\phi_\mu
	\label{diffsymm}
	\ee
in the particular case
	\be
	\phi_\mu \propto \partial_\mu \phi\ ,
	\ee
which defines the longitudinal diffeomorphisms \cite{Dalmazi:2020xou}. The most general 3D local integrated functionals invariant under \eqref{longsimm} are
	\begin{align}
	S_{\textsc{lg}}&=\int d^3x\ (\partial_\mu h\partial^\mu h- \partial_\rho h_{\mu\nu}\partial^\rho h^{\mu\nu}-2\partial_\mu h\partial_\nu h^{\mu\nu}+2\partial_\rho h_{\mu\nu}\partial^\mu h^{\nu\rho})\label{lgterm}\\
	S_{fr}&=\int d^3x\ (\partial_\rho h_{\mu\nu}\partial^\mu h^{\nu\rho}-\partial_\rho h_{\mu\nu}\partial^\rho h^{\mu\nu}) \label{fractterm}\\
	S_{m}&=\int d^3x\ \epsilon^{\mu\nu\rho} h_{\mu}^{\;\lambda}\partial_\nu h_{\rho\lambda}\label{massterm}\ ,
	\end{align}
where $h(x)\equiv \eta^{\mu\nu}h_{\mu\nu}(x)$ is the trace of the tensor field $h_{\mu\nu}(x)$. Assigning canonical mass dimension $[h_{\mu\nu}] = \frac{1}{2}$, we see that $S_m$ \eqref{massterm} is a lower dimensional term with respect to $S_{\textsc{lg}}$ \eqref{lgterm} and $S_{fr}$ \eqref{fractterm}.
Each of the above terms has an important physical meaning. $S_{\textsc{lg}}$ \eqref{lgterm} is the action of Linearized Gravity, which is invariant under the more restrictive infinitesimal diffeomorphism transformation \eqref{diffsymm}. The second term $S_{fr}$ \eqref{fractterm}, once matter is introduced, describes the dynamics of the fracton quasi-particles \cite{Pretko:2016lgv}, in its covariant extension described in \cite{Bertolini:2022ijb}. We will refer to \eqref{fractterm} as the ``fractonic" term of the invariant action $S_{inv}$. Finally, $S_{m}$ \eqref{massterm} is peculiar to 3D spacetime only, due to the presence of the Levi-Civita tensor $\epsilon^{\mu\nu\rho}$. It encodes a higher rank CS theory of fractons, whose Hall-like behaviour has been described in \cite{Bertolini:2024yur}. In \cite{Bertolini:2025rhz} it has been shown that $S_{m}$ \eqref{massterm} induces a mechanism of ``quasi topological'' mass generation for 3D Linearized Gravity. The aim of this paper is to study whether a similar mechanism of mass generation holds for the pure fracton theory, and the action we consider is 
	\be
	S_{inv}=S_{fr}+mS_m\ ,
	\label{Sinv}
	\ee
where $m$ is a massive parameter $[m]=1$.

\subsection{Gauge-fixing term}\label{subsec-gauge}

In order to identify the parameter $m$ appearing in the action $S_{inv}$ \eqref{Sinv} as a mass of the tensor field $h_{\mu\nu}(x)$, we have to show that it is a pole of the propagator $\langle h_{\mu\nu}(x)h_{\rho\sigma}(x')\rangle$. To compute the propagators of a gauge field theory we need a gauge-fixing condition, and the propagators are more easily computed if the gauge condition is covariant.
 Being the local gauge parameter $\phi(x)$ a scalar field, the most general gauge condition is
	\be
	\kappa_0\partial^\mu\partial^\nu h_{\mu\nu}+\kappa_1 \Box h = 0	\ ,
	\label{gaugefixing}
	\ee
where $\Box\equiv\partial^\mu\partial_\mu$ is the d'Alambert operator and $\kappa_0$ and $\kappa_1$ are dimensionless gauge parameters analogous to those appearing in the gauge fixed 4D Linearized Gravity \cite{Gambuti:2021meo,Gambuti:2020onb}. The gauge-fixing condition \eqref{gaugefixing} is implemented by the gauge-fixing term 
	\be
	S_{gf} = \int d^3x\ 
	b \left(
	\kappa_0\partial^\mu\partial^\nu h_{\mu\nu}+\kappa_1 \partial^2h + \frac{k}{2} b\right)\ ,
	\label{Sgfb}
	\ee
where $k$ is the usual gauge parameter, which is massive
	\be
	[k]=2\ ,
	\ee
and $b(x)$ is the Nakanishi-Lautrup Lagrange multiplier \cite{Nakanishi:1966zz,Lautrup:1967zz} for  the gauge  condition \eqref{gaugefixing}, with
	\be
	[b]=\frac{1}{2}\ .
	\ee
A massive gauge parameter induces infrared divergences, hence the Landau gauge choice is mandatory \cite{Alvarez-Gaume:1989ldl}
	\be
	k=0\ ,
	\label{Landaugauge}
	\ee
therefore, the total gauge-fixed action is
	\be
	S=\left.S_{inv}+S_{gf}\right|_{k=0}\ .
	\label{totactfract}
	\ee
Requiring that both $h_{\mu\nu}(x)$ and the transformed field $h'_{\mu\nu}(x)= h_{\mu\nu}(x)+\partial_\mu\partial_\nu \phi(x)$ satisfy the gauge condition \eqref{gaugefixing}, yields
	\be
	 (\kappa_0+\kappa_1)\Box^2 \phi=0 \label{residual_phi}\ .
	 \ee
Hence, in order to satisfy equation \eqref{residual_phi}, one must either have
	\be
	\kappa_0+\kappa_1 = 0\label{eq_kappa}
	\ee
or
	\be
	\Box^2 \phi=0\label{eq_phi}\ .
	\ee
A particular solution of the polyharmonic gauge condition \eqref{eq_phi} is the harmonic one,
	\be
	\Box \phi=0\ ,
	\ee
which is the usual residual gauge condition of abelian gauge theories \cite{Itzykson:1980rh}. The choice \eqref{eq_phi} turns out to be the only possibility in order to have the propagators, since, as we will see, \eqref{eq_kappa} corresponds to a pole of the propagators.

\section{Propagator}\label{propagators-g=0}

Among the propagators of the theory, computed in Appendix \ref{app-prop}, the tensor field propagator in moment space is
	\begin{align} 
	 \tilde{\Delta}_{\alpha\beta,\rho\sigma}(p)&=\frac{1}{2(p^2 +m^2)}\left[-2A^{(0)}-A^{(1)}-A^{(2)}+A^{(3)}+7A^{(4)}+\tfrac{m}{2p^2}\left(A^{(5)}+3A^{(6)}\right)\right]_{\alpha\beta,\rho\sigma}+\quad \nonumber \\
	 &\quad+\frac{1}{2p^2}\left[\tfrac{\kappa_0+3\kappa_1}{\kappa_0+\kappa_1}A^{(2)}-A^{(3)}-\tfrac{(\kappa_0+3\kappa_1)^2}{(\kappa_0+\kappa_1)^2}A^{(4)}\right]_{\alpha\beta,\rho\sigma}\label{prop-hh-frac}\ ,
	\end{align}
where the $\{A^{(i)}_{\alpha\beta,\rho\sigma}(p)\}$ form the following basis (see Appendix \ref{app-prop} for details)
\bea
	A^{(0)}_{\alpha\beta,\rho\sigma}  &=& \frac{1}{2}(\eta_{\alpha\rho}\eta_{\beta\sigma}+\eta_{\alpha\sigma}\eta_{\beta\rho}) 
	\label{A0}\\
	A^{(1)}_{\alpha\beta,\rho\sigma} &=& 
	\frac{1}{p^2}(\eta_{\alpha\rho}p_\beta p_\sigma +  
	\eta_{\alpha\sigma}p_\beta p_\rho +
	\eta_{\beta\rho}p_\alpha p_\sigma +
	\eta_{\beta\sigma}p_\alpha p_\rho)
	\label{A1}\\
	A^{(2)}_{\alpha\beta,\rho\sigma} &=& 
	\frac{1}{p^2}(\eta_{\alpha\beta} p_\rho p_\sigma +
	\eta_{\rho\sigma} p_\alpha p_\beta)
	\label{A2} \\
	A^{(3)}_{\alpha\beta,\rho\sigma} &=& 
	\eta_{\alpha\beta}\eta_{\rho\sigma}
	\label{A3}\\
	A^{(4)}_{\alpha\beta,\rho\sigma} &=& 
	\frac{p_\alpha p_\beta p_\rho p_\sigma}{p^4}
	\label{A4}\\
	A^{(5)}_{\alpha\beta,\rho\sigma} &=& 
	ip^\lambda (
	\epsilon_{\alpha\lambda\rho} \eta_{\sigma\beta} +
	\epsilon_{\beta\lambda\rho} \eta_{\sigma\alpha} +
	\epsilon_{\alpha\lambda\sigma} \eta_{\rho\beta} +
	\epsilon_{\beta\lambda\sigma} \eta_{\rho\alpha}
	)
	\label{A5}\\
	A^{(6)}_{\alpha\beta,\rho\sigma} &=&
	 \frac{ip^\lambda}{p^2} (
	\epsilon_{\alpha\lambda\rho} p_\sigma p_\beta +
	\epsilon_{\alpha\lambda\sigma} p_\rho p_\beta +
	\epsilon_{\beta\lambda\rho} p_\sigma p_\alpha +
	\epsilon_{\beta\lambda\sigma} p_\rho p_\alpha
	)\ .
	\label{A6}
	\eea
The propagator \eqref{prop-hh-frac} displays a massive pole
	\be
	 p^2 = -m^2 \ ,
	\label{poleg2}
	\ee
and has a good massless limit
	\be
	 \tilde{\Delta}_{\alpha\beta,\rho\sigma}(p)=\frac{1}{2p^2}\left[-2A^{(0)}-A^{(1)}+\tfrac{2\kappa_1}{\kappa_0+\kappa_1}A^{(2)}+2\tfrac{3\kappa_0^2+4\kappa_1\kappa_0-\kappa_1^2}{(\kappa_0+ \kappa_1)^2}A^{(4)}\right]_{\alpha\beta,\rho\sigma}\label{prop-hh-frac-massless}\ ,
	\ee
which agrees with \cite{Blasi:2022mbl}. As anticipated, we notice a pole at 
\be
\kappa_0+\kappa_1=0\ ,
\label{}\ee
which we take care of by keeping away from it in the gauge-fixing condition \eqref{gaugefixing}.  Having a well defined massless limit, the action \eqref{Sinv} describes a massive theory for the higher rank tensor field $h_{\mu\nu}(x)$, where the mass is introduced by means of a mechanism which closely reminds the one which characterizes the MCS theory \cite{Deser:1981wh,Deser:1982vy}. The analogy is even more evident if we introduce the fracton field strength \cite{Bertolini:2022ijb}
	\be
	F_{\mu\nu\rho}\equiv\partial_\mu h_{\nu\rho}+\partial_\nu h_{\rho\mu}-2\partial_\rho h_{\mu\nu}\ ,\label{field_strength}
	\ee
which is invariant under the generalized gauge transformation \eqref{longsimm}
\begin{equation}
    \delta F_{\mu\nu\rho}=0\ ,
\end{equation}
and satisfies the cyclic property
	\be\label{cycl}
	F_{\mu\nu\rho}+F_{\nu\rho\mu}+F_{\rho\mu\nu}=0\ ,
	\ee
and the Bianchi-like identity
	\begin{equation}\label{Bianchi 3D fractons}
	\epsilon_{\mu\nu\rho}\partial^\mu F^{\alpha\nu\rho}=0\ .
	\end{equation}
The action $S_{inv}$ \eqref{Sinv} can be written in terms of the fracton field strength \eqref{field_strength} as follows
	\be\label{Sinv(F)}
	S_{inv}=\int d^3x\left(-\frac1 6 F_{\mu\nu\rho}F^{\mu\nu\rho}+m\epsilon^{\mu\nu\rho}h_\mu^{\ \lambda}\partial_\nu h_{\rho\lambda}\right)\ ,
	\ee
which closely reminds the MCS theory
	\be
	S_{\textsc{mcs}}=\int d^3x\left(-\frac1 4 F_{\mu\nu}F^{\mu\nu}+\frac M 2\epsilon^{\mu\nu\rho}A_\mu\partial_\nu A_\rho\right)\ .
	\label{SMCS}\ee
The action \eqref{Sinv(F)} provides a 3D mass generation mechanism for the ``scalar charge theory of fractons'' discussed in \cite{Pretko:2016kxt,Pretko:2016lgv}.

\section{Degrees of freedom}

In order to analyze the number and nature of the DoF in the theory defined by the
action $S_{inv}$ \eqref{Sinv}, we begin by considering the MCS case as a preliminary example. This serves both as a
warm-up for the method we will adopt and as a physically relevant model in its own right.

\subsection{Maxwell-Chern-Simons}

The EoM of the MCS action $S_{\textsc{mcs}}$ \eqref{SMCS} are~:
\be
\frac{\delta S_{\textsc{mcs}}}{\delta A_\alpha} = -\partial_\mu F^{\alpha\mu} + M\epsilon^{\alpha\mu\nu}\partial_\mu A_\nu\ .
\label{eomSMCS}\ee
By separating time and space components, and working in the Coulomb gauge
\be
\partial^a A_a = 0\ ,
\label{coulombgauge}\ee
we obtain the following on-shell equations~:
\bea
\frac{\delta S_{\textsc{mcs}}}{\delta A_0} &=& \nabla^2 A^0 + M \epsilon^{0ab}\partial_aA_b=0\label{eomA0}\\
\frac{\delta S_{\textsc{mcs}}}{\delta A_a} &=& \Box A^a - \partial^a\partial_0A^0 
+ M \epsilon^{0ab}(\partial_bA_0-\partial_0A_b)=0 \ ,\label{eomAa}
\eea
where $\nabla^2\equiv\partial_a\partial^a$ is the Laplace operator. Let us first consider the massless case, $i.e.$ the pure Maxwell theory in 3D~:
\begin{description}
\item [Massless case $\mathbf{M=0}$]:
from \eqref{eomA0} we obtain
\be
\nabla^2 A_0=0\ ,
\label{poissonA0}\ee
a Poisson equation whose solution is 
\be
A_0=0
\label{A0=0}\ee
for fields vanishing at spatial infinity. Substituting this result into \eqref{eomAa}, gives
\be
\Box A_a=0\ .
\label{boxAa=0}\ee
Together with the transversality condition \eqref{coulombgauge}, this describes a single massless propagating DoF
corresponding to a planar transverse spin-1 mode, 
thus recovering the well knowns fact that in 3D the photon is equivalent to a massless scalar field, known as the ``dual photon'' \cite{Turner:2019wnh}.

\item[Massive case $\mathbf{M\neq0}$]: we decompose $A_a(x)$ into its transverse and longitudinal components
\be
A_a=u_a+\partial_a\psi\ ,
\label{decompA}\ee
with
\be
\partial^au_a=0\quad\Rightarrow\quad u_a=\epsilon_{0ab}\partial^bu\ ,
\label{ua}\ee
where $u(x)$ and $\psi(x)$ are two scalar fields. The Coulomb gauge \eqref{coulombgauge} then implies
\be
\nabla^2\psi=0\quad\Rightarrow\quad\psi=0\ .
\label{psi=0}\ee
The on-shell EoM \eqref{eomA0}, on the gauge condition \eqref{psi=0}, yields the Poisson equation
\be
\nabla^2(A^0+Mu)=0\ ,
\ee
whose solution relates the time component $A_0(x)$ to the longitudinal component of $A_a(x)$ 
\be
A_0=Mu\ ,
\label{A0=Mu}\ee
which generalizes \eqref{A0=0} to the massive case. Using \eqref{psi=0} and \eqref{A0=Mu} into the EoM \eqref{eomAa}, we get
\be
(\Box-M^2)A_a=0\ ,
\label{KGAa}\ee
which generalizes the massless wave equation \eqref{boxAa=0} to a Klein-Gordon equation for $A_a(x)$ which shows that the MCS theory propagates a single massive scalar mode, since $A_a(x)$ is a planar transverse vector.
\end{description}
We therefore recovered that both 3D Maxwell and MCS theories display one propagating DoF, as expected since the topological nature of the CS term cannot change the number of DoF.

\subsection{Massive fractons}

To count the DoF of the theory, and in order to make contact with a well-established procedure adopted in similar cases, we follow the approach of Deser, Jackiw, and Templeton for 3D Linearized Gravity, based on the decomposition of the rank-2 symmetric tensor field $h_{\mu\nu}(x)$. This approach can be found in \cite{Deser:1981wh,Deser:1982vy}, and, for 4D Linearized Gravity, in \cite{Carroll:2004st}. The EoM of the action $S_{inv}$ \eqref{Sinv} are
\be
\frac{\delta S_{inv}}{\delta h_{\alpha\beta}} =
-\partial_\mu F^{\alpha\beta\mu} +
m(\epsilon^{\alpha\mu\nu}\partial_\mu h_\nu^{\ \beta} + 
\epsilon^{\beta\mu\nu}\partial_\mu h_\nu^{\ \alpha} )\ .
\label{eomSinv}\ee
Following \cite{Deser:1981wh,Deser:1982vy}, we decompose the spacetime components of the symmetric tensor field $h_{\mu\nu}(x)$ into scalar, transverse, longitudinal and, for $h_{ab}(x)$, solenoidal parts
\bea
h_{00} &=& \lambda \label{h00}\\
h_{0a} &=& u_a + \partial_a\psi \label{h0a} \\
h_{ab} &=& (\eta_{ab}-\hat\partial_a\hat\partial_b)\varphi
+\hat\partial_a\hat\partial_b\chi
+\partial_a\phi_b + \partial_b\phi_a \label{hab}\ ,
\eea
where $\hat\partial_a\equiv\frac{\partial_a}{\sqrt{\nabla^2}}$, and $u_a(x)$ and $\phi_a(x)$ are transverse planar vectors
\be
\partial^au_a=\partial^a\phi_a=0\ ,
\label{transverseuxi}\ee
hence
\be
u_a=\epsilon_{0ab}\partial^bu\quad ; \quad \phi_a=\epsilon_{0ab}\partial^b\xi\ ,
\label{defuxi}\ee
so that the six components of $h_{\mu\nu}(x)$ are parametrized as follows
\be
h_{\mu\nu}=
\{\lambda\; ;\;u\; ;\; \psi\; ;\; \varphi\; ;\; \chi\; ;\; \xi\}\ .
\label{6components}\ee
From \eqref{eomSinv} and going on-shell, we have
\be
\frac{\delta S_{inv}}{\delta h_{00}}=0\quad\Rightarrow\quad\nabla^2(\lambda-\partial_0\psi-mu)=0\ ,
\label{eomh00}\ee
which gives
\be
\lambda=\partial_0\psi+mu\ .
\label{eom1}\ee
From
\be
\frac{\delta S_{inv}}{\delta h_{0a}}=0\ ,
\label{eomh0a}\ee
and using \eqref{eom1}, we get
\be
\epsilon^{0ab}\partial_b
[ (\Box-m^2+\nabla^2)u - \partial_0\nabla^2\xi-m\varphi]
+ \partial^a(-\nabla^2\psi+\partial_0\chi+m\nabla^2\xi)=0\ .
\label{eom2.0}\ee
Taking the divergence $\partial_a\eqref{eom2.0}$, we find the Poisson equation
\be
\nabla^2[\partial_0\chi+\nabla^2(m\xi-\psi)]=0\quad\Rightarrow \quad
\partial_0\chi+\nabla^2(m\xi-\psi)=0\ .
\label{eom2}\ee
Moreover, acting with $\epsilon_{0ka}\partial^k\eqref{eom2.0}$ we find, again, a Poisson equation which gives
\be
(\Box-m^2)u+\nabla^2(u-\partial_0\xi)-m\varphi=0\ .
\label{eom3}\ee
Finally, from
\be
\frac{\delta S_{inv}}{\delta h_{ab}}=0\ ,
\label{eomhab}\ee
and using \eqref{eom2}, we get
\bea
[2(\eta^{ab}-\hat\partial^a\hat\partial^b)\Box
+m(\epsilon^{0am}\hat\partial^b + \epsilon^{0bm}\hat\partial^a)\partial_0\hat\partial_m]\varphi &&
 \nonumber\\
 +[(\epsilon^{0am}\partial^b+\epsilon^{0bm}\partial^a)\partial_m
 (2\partial_0^2-\nabla^2+m^2)
 +2m(\eta^{ab}\nabla^2-\partial^a\partial^b)\partial_0
 ]\xi&&
 \nonumber\\
 -[(\epsilon^{0am}\partial^b+\epsilon^{0bm}\partial^a)\partial_m\partial_0
 +2m(\eta^{ab}\nabla^2-\partial^a\partial^b)
 ]u &=&0\ ,\label{eom4.0}
\eea
whose trace gives
\be
\Box\varphi + m\nabla^2(\partial_0\xi-u)=0\ ,
\label{eom4}\ee
and by computing $\epsilon_{0ac}\partial^c\partial_b\eqref{eom4.0}$ we find 
\be
m\partial_0\varphi - (\Box-m^2)\nabla^2\xi - \nabla^2\partial_0(u-\partial_0\xi)=0\ .
\label{5}\ee
In order to have the relevant equations at hand, we summarize them altogether~:
\bea
\lambda - \partial_0\psi-mu &=& 0 \label{eom1summ} \\
\partial_0\chi+\nabla^2(m\xi-\psi) &=& 0 \label{eom2summ} \\
(\Box-m^2)u+\nabla^2(u-\partial_0\xi)-m\varphi &=& 0 \label{eom3summ} \\
\Box\varphi + m\nabla^2(\partial_0\xi-u) &=& 0 \label{eom4summ} \\
m\partial_0\varphi - (\Box-m^2)\nabla^2\xi - \nabla^2\partial_0(u-\partial_0\xi) &=& 0 \label{eom5summ}\ .
\eea
The number and the nature of the DoF -- massive or massless, propagating or not -- are gauge independent. However, 
particular gauge choices can help make these aspects manifest. Due to the symmetry  \eqref{longsimm}, which involves the scalar gauge parameter $\phi(x)$, a scalar gauge-fixing condition is required. A convenient gauge choice to exhibit the structure of the theory  is 
\be
u=0\ ,
\label{u=0}\ee
which, from \eqref{h0a} and  \eqref{defuxi} implies that $h_{0a}(x)$ is a longitudinal planar vector 
\be
h_{0a}=\partial_a\psi\ ,
\label{h0a=0}\ee
and simplifies the identification of the propagating modes. Moreover, as we will see, this choice has the physical significance of allowing the fractonic sector of the theory to be identified. In this gauge, Eqs. \eqref{eom1summ}-\eqref{eom5summ} reduce to 
\bea
\lambda - \partial_0\psi &=& 0 \label{eom1gauge} \\
\partial_0\chi+\nabla^2(m\xi-\psi) &=& 0 \label{eom2gauge} \\
\nabla^2\partial_0\xi+m\varphi &=& 0 \label{eom3gauge} \\
\Box\varphi + m\nabla^2\partial_0\xi &=& 0 \label{eom4gauge} \\
m\partial_0\varphi - (\Box-m^2)\nabla^2\xi + \nabla^2\partial_0^2\xi &=& 0 \label{eom5gauge}\ .
\eea

\begin{description}
\item [Massless case $\mathbf{m=0}$:]
let us consider first the massless case, where the above equations become
\bea
\lambda - \partial_0\psi &=& 0 \label{eom1m=0} \\
\partial_0\chi-\nabla^2\psi &=& 0 \label{eom2m=0} \\
\nabla^2\partial_0\xi &=& 0 \label{eom3m=0} \\
\Box\varphi &=& 0 \label{eom4m=0} \\
\Box\nabla^2\xi - \nabla^2\partial_0^2\xi &=& 0 \label{eom5m=0}\ ,
\eea
which imply that the massless theory described by the pure fractonic action $S_{fr}$ \eqref{fractterm}, which corresponds to the Maxwell action, features two DoF: $\varphi(x)$, associated to the transverse component of $h_{ab}(x)$ through \eqref{hab} and satisfying the wave equation \eqref{eom4m=0}, and $\chi(x)$, associated with the longitudinal sector. The fields $\psi(x)$ and $\lambda(x)$ are determined by $\chi(x)$ through \eqref{eom2m=0} and \eqref{eom1m=0}, while $\xi(x)$, associated to the solenoidal part of $h_{ij}(x)$, due to \eqref{eom3m=0} is constant
\be
\partial_0\xi=0\ ,
\ee
as a consequence of which \eqref{eom5m=0} becomes $(\nabla^2)^2\xi=0$, which is solved by
\be
\xi=0\ .
\label{xi=0}\ee

\item [Massive case $\mathbf{m\neq0}$:]
in the longitudinal gauge \eqref{u=0}, described by Eqs. \eqref{eom1gauge}-\eqref{eom5gauge}, from \eqref{eom3gauge} and \eqref{eom4gauge} we immediately find that $\varphi(x)$ satisfies the Klein-Gordon equation
\be
(\Box-m^2)\varphi=0\ .
\label{kgphi}\ee
Alternatively, using \eqref{eom3gauge} in \eqref{eom5gauge} we find that $\nabla^2\xi(x)$ satisfies the same Klein-Gordon equation 
\be
(\Box-m^2)\nabla^2\xi=0\ ,
\label{kgxi}\ee
but $\nabla^2\xi(x)$ and $\varphi(x)$ are non independent one from each other because they are related by \eqref{eom3gauge}.
Hence, $\varphi(x)$ -- or, equivalently, $\nabla^2\xi(x)$ -- is a propagating massive DoF. The field $\chi(x)$ remains dynamical and continues to determine $\psi(x)$ and $\lambda(x)$, as in the massless case.
\end{description}
Summarizing, the symmetric rank-2 tensor field $h_{\mu\nu}(x)$ contains six scalar components \eqref{6components}. In the massless theory defined by the action $S_{fr}$ \eqref{fractterm}, two of these -- $\varphi(x)$ and $\chi(x)$ -- correspond to independent DoF. 
The presence of the lower dimensional CS-like term $S_m$ \eqref{massterm} does not change the number of propagating DoF, as in MCS theory.
In fact, the number of DoF, either in the massless or in the massive case is thus two. The difference is that, in the massive case, one of the DoF satisfies a Klein-Gordon equation, hence it is a propagating massive one. It is interesting to notice that the CS-like term \eqref{massterm}, although not topological because metric dependent, behaves as the topological CS term in MCS, namely gives mass without changing the number of DoF, as expected for a topological term. This confirms the ``quasi'' topological nature of the CS-like term \eqref{massterm}, already remarked in \cite{Bertolini:2025qcy}.
It is also worth noting that the two DoF of the theory -- $\nabla^2\xi(x)$ and $\chi(x)$  -- belong to the spatial component $h_{ab}(x)$ of the tensor field $h_{\mu\nu}(x)$, as it happens also in 4D Linearized Gravity \cite{Carroll:2004st, Weinberg:1995mt}. Hence, one might argue that, by adopting -- as in 4D Linearized Gravity -- a vector gauge fixing, in particular the condition
\be
h_{0\mu}=0\ \Leftrightarrow\ \lambda=u=\psi=0\ ,
\label{vectorgf}\ee
instead of the scalar one \eqref{gaugefixing}, we would obtain the same physical content. This is despite the fact that a vector gauge condition like \eqref{vectorgf} corresponds to the infinitesimal diffeomorphism transformation \eqref{diffsymm}, which is not a symmetry of the action $S_{fr}$ \eqref{fractterm}. Nonetheless, such a situation may arise in fractonic theories. Indeed, in \cite{Bertolini:2023juh}, where fractons coupled to 4D Linearized Gravity were studied, it was shown that this property holds: over-gauging the longitudinal diffeomorphisms \eqref{longsimm} with a vector gauge condition, instead of the more appropriate scalar one, leads to the same number of DoF.
A closer inspection here, however, shows that this is not the case\footnote{We thank our Referee for this remark}. In fact, considering the gauge fixing \eqref{vectorgf}, Eq. \eqref{eom2gauge} reduces to
\be
\partial_0\chi+m\nabla^2\xi=0\ ,
\ee
which relates the two DoF previously identified. Hence, a vectorial gauge fixing is not applicable in this context, since over-gauging the theory with a vector condition removes one DoF out of two. This is perfectly natural -- if anything, the opposite would have been surprising.

\section{Physical interpretation: generalized electromagnetism and fractons}

In analogy to the ordinary MCS theory \cite{Deser:1981wh, Deser:1982vy,Dunne:1998qy} and 4D covariant fractons \cite{Bertolini:2022ijb}, we define the generalized electromagnetic fields in terms of the invariant fracton field strength \eqref{field_strength} as
	\begin{align}
	E^{ij} &\equiv F^{ij0}=\partial^ih^{j0}+\partial^jh^{i0}-2\partial^0h^{ij} \label{def_E}\\
	B^i &\equiv -\frac{2}{3} \epsilon_{0jk} F^{ijk}=-2\epsilon_{0jk}\partial^j h^{ik}\ ,    \label{def_B}
	\end{align}
which gives
	\begin{equation}
	F^{ijk} = -\frac{1}{2}\left(\epsilon^{0ki}B^j + \epsilon^{0kj}B^i\right)\ .
	\end{equation}
We can combine \eqref{h00}, \eqref{h0a} and \eqref{eom1summ} in
	\be
	h_{\mu0}=\partial_\mu\psi+\epsilon_{0\mu\nu}\partial^\nu u-m\eta_{\mu0}u\ ,
	\ee
which, in the longitudinal gauge \eqref{u=0}, becomes
\begin{equation}
    h_{\mu0} = \partial_\mu\psi \label{particular_solution}\ .
\end{equation}
Therefore the longitudinal gauge choice \eqref{u=0} implies \eqref{particular_solution}, which coincides with the solution in \cite{Bertolini:2022ijb}. In this sense, as anticipated, that gauge choice enhances the fractonic interpretation of the theory.
In this case, due to the cyclic property \eqref{cycl}, the field strength \eqref{field_strength} satisfies also \cite{Bertolini:2022ijb}
	\begin{align}
	F^{i00}&= 0   \label{Fi00=0}\\ 
	F^{i0j} &= -\frac{1}{2}F^{ij0}\ , \label{Fi0j}
	\end{align}
and we recover the same expression of the fractonic electric tensor field given in \cite{Pretko:2016lgv,Bertolini:2022ijb}
	\be
	E_{ij} = 2\left(\partial_0 h_{ij} - \partial_i \partial_j \psi\right)\ ,
	\label{Eij}\ee
which is a symmetric rank-2 generalization of the expression of the electric field of the ordinary $U(1)$ vector gauge theory. 
In terms of the generalized electric \eqref{Eij} and magnetic \eqref{def_B} fields, the EoM \eqref{eomh0a} can be written
\be\label{Gauss-vec}
	\partial_j E^{ij}=-m B^i \equiv  -\rho^i_{\textsc{cs}} \ ,
\ee
from which it also follows that
	\begin{equation}
	\partial_i\partial_j E^{ij}=-m\partial_i B^i\equiv\rho_{\textsc{cs}} \label{Gauss_vacuum}\ , 
	\end{equation}
which are fractonic Gauss constraints \cite{Pretko:2016kxt,Pretko:2016lgv,Bertolini:2024jen} where  the magnetic field $B^i(x)$ \eqref{def_B} plays the role of an internal ``electric'' charge related to the CS-like term \eqref{massterm} in the action $S_{inv}$ \eqref{Sinv}. In particular, $\rho^i_{\textsc{cs}}(x)$ can be seen as a dipole density through
	\be\label{D}
	D^i_{\textsc{cs}}\equiv\int d\Sigma\, x^i\rho_{\textsc{cs}}=-m\int d\Sigma\, x^i\partial_jB^j=\int d\Sigma\, \rho^i_{\textsc{cs}}\ ,
	\ee
with $ d \Sigma \equiv  d x_1  d x_2$ and $D^i_{\textsc{cs}}(t)$ the total dipole moment. Therefore, the CS-like term \eqref{massterm} plays the role  of ``intrinsic'' matter, in the following sense. In ``standard'' fracton theories, the fractonic charge and current densities $\rho(x)$ and $J^{ij}(x)$ depend on external matter fields. In our case, however, the right-hand side of the Gauss constraint \eqref{Gauss_vacuum}, namely $\rho_{\textsc{cs}}(x)$, and $J^{ij}_{\textsc{cs}}(x)$, which will appear shortly at the right-hand side of the Amp\`ere-like equation \eqref{Ampere_vacuum}, are expressed directly in terms of the tensor field $h_{\mu\nu}(x)$. In particular, in \eqref{Gauss_vacuum} the quantity $\rho_{\textsc{cs}}(x)$ is given by the divergence of the generalized magnetic field $B^i(x)$, itself written in terms of the tensor field in \eqref{def_B}. Similarly, $J^{ij}_{\textsc{cs}}(x)$ \eqref{J} depends on the generalized electric field, which is expressed in terms of $h_{ij}(x)$ in \eqref{Eij}.
We thus see that, due to the presence of the massive CS-like term, the tensor gauge field plays a dual role: that of matter, through $\rho_{\textsc{cs}}(x)$ and 
$J^{ij}_{\textsc{cs}}(x)$, and that of generalized electromagnetic fields, as in \eqref{def_E} and \eqref{def_B}. We refer to this matter as ``intrinsic'', in the sense that it originates from the tensor gauge field $h_{\mu\nu}(x)$ itself, rather than from external matter fields, as in QED, for example.
It is worth noting that the existence of such intrinsic matter is not peculiar to our theory.
The same property also appears in various frameworks of ordinary, non-fractonic, theories, such as those involving axions and $\theta$-terms. In particular, see \cite{Sikivie:1983ip,Wilczek:1987mv} (standard axions), \cite{Rosenberg:2010ia} (Witten effect in topological insulators), and \cite{Chatzistavrakidis:2020wum} (axions and $\theta$-terms in Linearized Gravity). Regarding fractons, in \cite{Pretko:2017xar,Bertolini:2022ijb,Bertolini:2023sqa} the same phenomenon has been observed  in the context of the Witten effect for higher-spin fracton theories. In that case, a quasitopological $\theta$-term is added to the pure fracton action. This suggests that the emergence of intrinsic matter is generally tied to the presence of topological or quasitopological terms.\\

The Gauss law \eqref{Gauss_vacuum} is crucial in fractonic theories, since it encodes the main constraint on the motion of these quasiparticles. In fact, by using it in the definition of total dipole moment $D^i_{\textsc{cs}}(t)$ \eqref{D} we have
	\be\label{D=0}
	D^i_{\textsc{cs}}=\int d\Sigma\, x^i\rho_{\textsc{cs}}=\int d\Sigma\, x^i\partial_a\partial_bE^{ab}=-\int d\Sigma\partial_aE^{ai}=0
	\ee
up to boundary terms. This means that the total dipole moment vanishes on the 2D surface. Hence, single charges must not move, since otherwise their motion would change the total dipole. We remark that the notion of charge neutrality here is the same as in ordinary electromagnetism, and it does not imply the absence of single electric charges. As discussed in \cite{Pretko:2016kxt}, charge neutrality does not conflict with the immobility property. For instance, consider the simplest ``neutral'' configuration: a dipole made of two opposite charges. If the system is constrained by conservation of the total dipole moment, as in \eqref{D=0}, then the individual charges cannot move independently -- as if they were isolated -- but only in such a way that the total dipole moment is preserved. In other words, it is the dipole as a whole that moves. Something similar also happens to the dipole-like vector density $\rho^i_{\textsc{cs}}(x)$. Notice that from the definition of the vector charge $\rho^i_{\textsc{cs}}(x)$ \eqref{Gauss-vec} and \eqref{D}, and of the magnetic field $B^i(x)$ \eqref{def_B}, it follows that the trace ${D_{\textsc{cs}}}^i_{\ i}(t)$ of the total quadrupole moment\footnote{We adopt the traceful definition of quadrupole moment, instead of the traceless one,  as in \cite{Pretko:2016lgv,Gromov:2018nbv} and in \cite{raab}.}
	\be\label{Dij}
	D^{ij}_{\textsc{cs}}\equiv\int d\Sigma\, x^ix^j\rho_{\textsc{cs}}
	\ee
vanishes. In fact
	\begin{equation}
	{D_{\textsc{cs}}}^i_{\ i}=\eta_{ij}D^{ij}_{\textsc{cs}}=\int  d\Sigma\  x^2 \rho_{\textsc{cs}}=2\int  d\Sigma\  x_i \rho^i_{\textsc{cs}}\propto \int  d\Sigma\  x_i B^i =0 \label{int_xdotB=0}\,,
	\end{equation}
which is a fractonic conservation \cite{Pretko:2016lgv}, in the same sense as \eqref{D=0}, since it constrains the motion of the internal dipole-like charge $\rho^i_{\textsc{cs}}(x)$ in the direction perpendicular to the vector $\rho^i_{\textsc{cs}}(x)$ itself \cite{Doshi:2020jso}.  It is interesting to observe that the theory described by the action $S_{inv}$ \eqref{Sinv} is the first case in which a fractonic behaviour is already present without introducing matter, as a consequence of the presence of the massive CS-like term, which therefore, besides giving a mass to the tensor field $h_{\mu\nu}(x)$, plays also he role of an ``intrinsic'' matter contribution.\\

It is easy to check that the EoM \eqref{eomhab}, in terms of the generalized electric \eqref{Eij} and magnetic \eqref{def_B} fields, becomes
\be
	-\partial_0 E^{ij}+\frac{1}{2}\left(\epsilon^{0ki}\partial_k B^j + \epsilon^{0kj}\partial_k B^i\right)=J^{ij}_{\textsc{cs}}\label{Ampere_vacuum}\ ,
	\ee
where
	\be\label{J}
	J^{ij}_{\textsc{cs}}=\frac{1}{2}m \left(\epsilon^{0ik}\mixt{E}{j}{k}+\epsilon^{0jk}\mixt{E}{i}{k}\right)\ ,
	\ee
which is traceless
	\be\label{TJ=0}
	J^{\ \, i}_{\!\textsc{cs}\,i}=0\ .
	\ee
The EoM \eqref{Ampere_vacuum} is an Amp\`ere-like equation of fractonic type \cite{Pretko:2016lgv} with an internal current-like contribution $J^{ij}_{\textsc{cs}}(x)$ associated to the CS-like massive term \eqref{massterm} and depending on the electric field $E^{ij}(x)$ \eqref{def_E}.\\

Fractonic behaviours may emerge in two possible ways: through Gauss-like constraints on the motion of the dipole coming from charge neutrality conditions, like \eqref{Gauss-vec} and \eqref{Gauss_vacuum}, and/or through continuity equations for the matter contribution \cite{Pretko:2016lgv}. Indeed to further confirm the fractonic nature of the theory and the role of ``intrinsic matter'' played by the CS-like term, we observe that taking the double divergence $\partial_i\partial_j$ of the Amp\`ere-like equation \eqref{Ampere_vacuum} and using the Gauss constraint \eqref{Gauss_vacuum}, we obtain a fractonic continuity equation \cite{Pretko:2016lgv}  for the fractonic charge $\rho_{\textsc{cs}}(x)$ and its dipole-like current $J^{ij}_{\textsc{cs}}(x)$ \eqref{J}  
	\be\label{cont-dipole}
	\partial_0\rho_{\textsc{cs}}+\partial_i\partial_jJ^{ij}_{\textsc{cs}}=0\ ,
	\ee
associated to dipole conservation through
	\be\label{d0D=0}
	\partial_0D^i_{\textsc{cs}}=\int d\Sigma\,x^i\partial_0\rho_{\textsc{cs}}=\int d\Sigma\,\partial_jJ^{ij}_{\textsc{cs}}=0\ .
	\ee
The above continuity equation \eqref{cont-dipole} also encodes the conservation of the trace ${D_{\textsc{cs}}}^i_{\ i}(t)$ of the quadrupole moment $D^{ij}_{\textsc{cs}} (t)$ \eqref{Dij}, as a consequence of the tracelessness of the intrinsic current $J^{ij}_{\textsc{cs}}(x)$ \eqref{TJ=0}
	\be\label{d0TrQ=0}
	\partial_0\int d\Sigma\left(x^2\rho_{\textsc{cs}}\right)=\int d\Sigma\,x_k\partial_jJ^{kj}_{\textsc{cs}}=-\int d\Sigma\, J^{\ \,i}_{\!\textsc{cs}\, i}=0\ .
	\ee
Both conservations \eqref{d0D=0} and \eqref{d0TrQ=0} reflect the constraints implied by \eqref{D=0} and \eqref{int_xdotB=0}, $i.e.$ immobile fractonic charge $\rho_{\textsc{cs}}(x)$, and dipole-like $\rho^i_{\textsc{cs}}(x)$ constrained on a line. Notice also that while dipole conservation comes from both the Gauss constraint \eqref{Gauss_vacuum} used in \eqref{D=0}, \textit{and} the continuity equation \eqref{cont-dipole} in \eqref{d0D=0}, the conservation of the trace of the quadrupole moment is not related to a Gauss law, but is encoded only in the continuity equation \eqref{cont-dipole} and in the tracelessness of $J^{ij}_{\textsc{cs}}(x)$ \eqref{TJ=0}, and thus is a direct consequence of the presence of the CS-like term, since it reflects the fact that the CS-like term \eqref{massterm} does not depend on the trace of the gauge field $h_{\mu\nu}(x)$ \cite{Bertolini:2024yur}. Interestingly, this  has an effect on the electric tensor field $E^{ij}(x)$ \eqref{def_E}, which, through the Gauss law \eqref{Gauss-vec}, is constrained to be globally traceless on-shell. In fact, from \eqref{int_xdotB=0} and \eqref{Gauss-vec} we get
	\be
	0=\int  d\Sigma\  x_i \rho^i_{\textsc{cs}}=-\int  d\Sigma\  x_i\partial_jE^{ij}\quad\Rightarrow\quad\int d\Sigma\, E^i_{\ i}=0\ . \label{trace_E_cs}
	\ee
Finally, from the  Bianchi-like identity \eqref{Bianchi 3D fractons} we have
	\be
	\epsilon_{k0j}\partial^k F^{i0j} + \epsilon_{kj0}\partial^k F^{ij0} + \epsilon_{0jk}\partial^0 F^{ijk}=0 \ , \label{Bianchi_i}
	\ee
which, using the definitions \eqref{def_E} and \eqref{def_B} can be written as
	\be\label{Faraday from Bianchi}
	\partial_0 B^i +\epsilon_{0jk}\partial^j E^{ik}=0\ ,
	\ee
    which strongly reminds the Faraday law of ordinary electromagnetism, and coincides with the fractonic higher-rank Faraday law introduced in \cite{Pretko:2017kvd}. The CS-like term \eqref{massterm} induces the mechanism of mass generation for the tensor field $h_{\mu\nu}(x)$, as it appears from the presence of the massive pole $p^2=-m^2$ in the propagator \eqref{prop-hh-frac}. Besides this important property, we already remarked that \eqref{massterm} behaves like a kind of intrinsic matter, since it induces charges and currents at the r.h.s. of the fractonic Gauss constraints \eqref{Gauss-vec}, \eqref{Gauss_vacuum} and of the Amp\`ere-like equation \eqref{Ampere_vacuum}. Finally, the CS-like term \eqref{massterm} can be seen as a fractonic matter contribution \cite{Pretko:2016lgv,Pretko:2017xar}, since, on the fracton solution \eqref{particular_solution}, we can write
	\be\label{Sm}
	mS_m=-\frac{1}{2}\int d^3x\, h_{\mu\nu}J^{\mu\nu}_{\textsc{cs}}\ ,
	\ee
where we defined the ``intrinsic'' matter current
	\be \label{def_J_cs}
	J_{\textsc{cs}}^{\mu\nu}\equiv-\frac{m}{3} \left(\epsilon^{\mu\alpha\beta}F^\nu_{\ \alpha\beta}+\epsilon^{\nu\alpha\beta}F^\mu_{\ \alpha\beta}\right)\ ,
	\ee
whose spatial components are given by \eqref{J} and, considering the definitions of fractonic charges $\rho_{\textsc{cs}}(x)$ \eqref{Gauss_vacuum} and $\rho_{\textsc{cs}}^i(x)$ \eqref{Gauss-vec}, 
	\be
	J_{\textsc{cs}}^{00}=0\quad;\quad J_{\textsc{cs}}^{0i}=-\frac{1}{2}m B^i=-\frac1 2 \rho^i_{\textsc{cs}}\quad;\quad\partial_iJ_{\textsc{cs}}^{0i}=\frac1 2 \rho_{\textsc{cs}} \label{rho_cs}\ .
	\ee
So that the CS-like term \eqref{massterm} reads
\be
mS_m	=\frac1 2\int d^3x\;\left(\psi\rho_{\textsc{cs}}-h_{ij}J^{ij}_{\textsc{cs}}\right)\ ,
\label{}\ee
which can be seen as a fractonic matter coupling \cite{Pretko:2016lgv,Pretko:2017xar}. The continuity equation \eqref{cont-dipole}, in this picture, naturally emerges as a consequence of gauge invariance, since, asking that the action $S_m$ \eqref{Sm} is invariant under the gauge transformation \eqref{longsimm}, and taking into account that $\delta\psi=\partial_0\phi$, gives
\be
\delta S_m = - \int d^3x\; 
\phi \left(\partial_0\rho_{\textsc{cs}}+\partial_i\partial_j J^{ij}_{\textsc{cs}}\right)=0\ ,
\label{}\ee
which implies \eqref{cont-dipole}.\\

We conclude this Section with a further interpretation of the nontrivial role of the CS-like term \eqref{massterm}. The fractonic charge density $\rho_{\textsc{cs}}(x)$, related to the CS-like action \eqref{massterm} by the definition \eqref{Gauss_vacuum}, represents one of the two propagating massive DoF of the theory found in Section 4.2. In fact, $\rho_{\textsc{cs}}(x)$ can be associated to the solenoidal sector of $h_{ij}(x)$ \eqref{hab}
\be
	\rho_{\textsc{cs}}=-m\partial_iB^i=2m\partial_i\left(\epsilon_{0jk}\partial^jh^{ik}\right)=-2m\nabla^2\xi\ ,
\ee
which, as we saw in Section 4.2, satisfies the Klein-Gordon equation \eqref{kgxi}, which therefore can be equivalently written as
\be\label{massiveKG}
	\left(\Box-m^2\right)\rho_{\textsc{cs}}=0\ .
	\ee
It may be useful to comment on the fact that a fracton quasiparticle can propagate, $i.e.$ it possesses a nonvanishing propagator. This is not contradictory, regardless of whether the fracton is massless or massive. In \cite{Bertolini:2022ijb}, the covariant extension of ordinary fracton theory was formulated as a gauge theory of a symmetric tensor field, and the propagators were explicitly computed, turning out to be nonvanishing.
According to the classical references on fractons, such as \cite{Pretko:2020cko, Pretko:2016kxt, Pretko:2016lgv, Vijay:2016phm, Vijay:2015mka}, a fracton can be defined within the scalar-charge tensor gauge theory through the generalized Gauss law
\begin{equation}
\partial_i \partial_j E^{ij} = \rho ,
\label{gauss}\end{equation}
which enforces both charge and dipole-moment conservation. These constraints can also be expressed through the higher-rank continuity equation
\begin{equation}
\partial_0 \rho + \partial_i \partial_j J^{ij} = 0 ,
\label{continuity}\end{equation}
which implies that isolated charges are immobile. In a fracton dipolar system, however, pairs of fractons can move (and thus propagate), provided the total dipole moment is preserved. In fact, the very term ``fracton'' originates from the idea of ``being a fraction of a mobile quasiparticle'' \cite{Vijay:2015mka}.
Our theory therefore provides two key insights:
\begin{enumerate}
\item the Klein-Gordon equation \eqref{kgphi} (or \eqref{kgxi} or \eqref{massiveKG}), which describes a massive, propagating fractonic DoF;
\item the Gauss \eqref{Gauss_vacuum} and continuity \eqref{cont-dipole} equations, which constrain this DoF to obey fractonic conservation laws, ensuring that its motion remains compatible with the conservation of the total dipole moment.
\end{enumerate}

\section{Energy-moment tensor}

Making explicit the metric dependence of the invariant action \eqref{Sinv(F)}
\begin{equation}
    S_{inv} = \int d^3 x\ \left(
    -\frac{1}{6}\sqrt{-g}\ g^{\mu\alpha}g^{\nu\beta}g^{\rho\gamma}F_{\alpha\beta\gamma}F_{\mu\nu\rho}+ m \epsilon^{\mu\nu\rho}g^{\lambda\sigma}h_{\mu\lambda} \partial_\nu h_{\rho\sigma}
    \right)\ ,
\end{equation}
we can compute the energy-moment tensor
\begin{align}
T_{\alpha\beta} &\equiv -\frac{2}{\sqrt{-g}}\fdv{S_{inv}}{g^{\alpha\beta}}\bigg|_{g=\eta}\nonumber\\
&= -\frac{1}{6}\eta_{\alpha\beta}F^2 +\frac{1}{3}\eta_{\alpha\gamma}\eta_{\beta\lambda}\left(2F^{\lambda\nu\rho}\mixt{F}{\gamma}{\nu\rho} + F^{\mu\nu\lambda}\invmixt{F}{\mu\nu}{\gamma}\right)-m \epsilon^{\mu\nu\rho}\left(h_{\mu\alpha}\partial_\nu h_{\rho\beta}+h_{\mu\beta}\partial_\nu h_{\rho\alpha} 
    \right)\ .\label{T}
\end{align}
On the particular solution \eqref{particular_solution} the components of \eqref{T} read
	\begin{align}
	T_{00}=& \frac{1}{4}\left( E^{ij} E_{ij}+B^i B_i \right)\label{T00}
	\\
	T_{0i}=& -\frac{1}{2}\left[\epsilon_{0ij}E^{jk}B_k +m\left(\partial_0 \psi B_i - \epsilon^{0jk}\partial_j\psi E_{ik}\right)\right]\label{T0i}\\
	T_{ij}=&\frac{1}{4}\eta_{ij}(E^{ab}E_{ab}+B^a B_a)- E_{i}^{\ a}E_{ja}+\frac{1}{2}B_i B_j +\label{Tij}\\
	&+\frac{m}{2} \left[\epsilon^{0mn}\left(h_{mi}E_{nj}+h_{mj}E_{ni}\right)-\left(\partial_i \psi B_j +\partial_j \psi B_i \right)\right]\ . \nonumber
	\end{align}
We observe that the energy density \eqref{T00} is positive definite and formally identical to the 3D electromagnetic one
	\begin{equation}
	u= \frac{1}{2}\left(E^i E_i + B^2\right)\ ,
	\end{equation}
of which it represents a higher rank extension.
We however notice that the CS-like contribution in $T_{\alpha\beta}(x)$ \eqref{T}  breaks gauge invariance by a term which is a total derivative 
	\be
	\delta T_{\alpha\beta}=-m\epsilon^{\mu\nu\rho}\partial_\mu\left(\partial_\alpha\phi\partial_\nu h_{\rho\beta}+\partial_\beta\phi\partial_\nu h_{\rho\alpha}\right)\ .
	\label{nongaugeinv}\ee
Notice that on the fractonic solution \eqref{particular_solution} one has
\be
\left.\delta T_{00}\right|_{\eqref{particular_solution}}=0\ ,
\label{}\ee
while $\left.\delta T_{\alpha i}\right|_{\eqref{particular_solution}}=\mbox{boundary terms}$. Hence in the fractonic embedding the energy density \eqref{T00} is indeed invariant.
The same issue of non-gauge invariance \eqref{nongaugeinv} affects Linearized Gravity \cite{Misner:1973prb,tong}, and is related to the fact that the action of the model is invariant up to boundary terms. Therefore, in order to obtain a gauge-invariant physical energy-moment tensor, an ``average'' over a length scale $l$ is performed, as done in Linearized Gravity \cite{Misner:1973prb,tong}: a convolution with a well behaved weighting function is taken over a spacetime volume of size $l$, in such a way that boundary terms can be neglected $\langle X\partial Y\rangle\sim-\langle Y\partial X\rangle$. In this way, the resulting ``averaged'' energy moment tensor
	\be
	\bar T_{\mu\nu}\equiv\langle T_{\mu\nu}\rangle \label{average_emt_def}
	\ee
 is  gauge invariant
	\be
	\delta\bar T_{\mu\nu}=0\ .
	\ee
In particular the averaged component $\bar T_{0i}(x)$ \eqref{T0i} reads
	\be
	\bar T_{0i}= -\frac{1}{2}\langle\epsilon_{0ij}E^{jk}B_k\rangle \label{<T0i>}\ ,
	\ee
where Faraday's law \eqref{Faraday from Bianchi} has been used, which
 strongly reminds the Poynting vector of 3D electromagnetism
	\begin{equation}
	S^i = \epsilon^{0ij}E_j B\ .
	\end{equation}
One might see also an analogy between $T_{ij}(x)$ \eqref{Tij} (and its average $\bar T_{ij}(x)$) and the electromagnetic stress tensor 
	\begin{equation}
	\sigma_{ij} = E_i E_j -\frac{1}{2}\eta_{ij}(E^i E_i +B^2)\ ,
	\end{equation}
since both are quadratic forms diagonal in the (generalized) electric and magnetic fields. The differences reside in the fact that the CS-like term \eqref{massterm} contributes to the energy moment tensor.
In the massless limit ($m\to0$) the contribution to the total energy moment tensor comes from the fractonic Maxwell-like action only $S_{fr}$ \eqref{fractterm}
	\be
	T_{\mu\nu}|_{m=0}=T_{\mu\nu}^{(\textit{fr})}\ ,
	\ee
which is gauge invariant without the need of the average prescription \eqref{average_emt_def}. The components of $T_{\mu\nu}^{(\textit{fr})}(x)$, on the solution \eqref{particular_solution}, read
	\begin{align}
	T^{(\textit{fr})}_{00}=& \frac{1}{4}\left( E^{ij} E_{ij}+B^i B_i \right)\label{T00fr}
	\\
	T^{(\textit{fr})}_{0i}=& -\frac{1}{2}\epsilon_{0ij}E^{jk}B_k \label{T0ifr}\\
	T^{(\textit{fr})}_{ij}=&\frac{1}{4}\eta_{ij}(E^{ab}E_{ab}+B^a B_a)- E_{i}^{\ a}E_{ja}+\frac{1}{2}B_i B_j \ . \label{Tijfr}
	\end{align}
Using the Gauss \eqref{Gauss_vacuum}, Amp\`ere \eqref{Ampere_vacuum}, and Faraday \eqref{Faraday from Bianchi} laws at $m=0$, we have
	\be\label{consT0}
	\left.\partial^{\alpha}T^{(\textit{fr})}_{\alpha0}\right|_{m=0}=\left(\partial^0T^{(\textit{fr})}_{00}+\partial^iT^{(\textit{fr})}_{i0}\right)_{m=0}=0\ ,
	\ee
which is the continuity equation for the energy density $T^{(\textit{fr})}_{00}(x)$  in the massless limit. On the other hand we have
	\be\label{breaking}
	\left.\partial^\alpha T^{(\textit{fr})}_{\alpha i}\right|_{m=0} =\frac{3}{4}B_i \partial^j B_j+\frac{1}{2} B_j\partial^j B_i  - \frac{1}{2}E^{jk}\partial_j E_{ik}\ ,
	\ee
    which therefore is not conserved, as in the 4D case \cite{Bertolini:2022ijb}\footnote{We remind that, as discussed in \cite{Bertolini:2022ijb}, the energy-moment tensor defined as \eqref{T} is the conserved current associated to the diffeomorphism invariance, and it should not be conserved in a theory invariant under \eqref{longsimm}, which is a subclass of the diffeomorphism transformation \eqref{diffsymm}.}. Nonetheless this result is useful to identify a fractonic force, which can be done as follows. 
In ordinary electromagnetism matter is introduced by adding a source term to the Maxwell action
	\be\label{Stot}
	S_{\textit{Max}}\ \to\ S_{\textit{Max}}+\int d^3x A_\mu J^\mu\ ,	
	\ee
which contributes to the total energy-moment tensor by means of an additional term $T_{\mu\nu}^{(J)}(x)$
\be
T_{\mu\nu}^{(tot)} = T_{\mu\nu}^{(\textit{Max})} + T_{\mu\nu}^{(J)}\ ,
\label{}\ee
where $T_{\mu\nu}^{(\textit{Max})}(x)$ is the energy-moment tensor associated to the Maxwell action $S_{\textit{Max}}$.
The Lorentz force can be identified as the contribution from $T_{\mu\nu}^{(J)}(x)$, as \cite{DEGROOT196877,Medina:2017mcd}
	\be\label{Tem}
f_\nu\equiv\partial^\mu T_{\mu\nu}^{(J)}\ ,
	\ee
so that, from the conservation of the total energy-moment tensor, we have
	\be\label{Fmax}
	f_\nu=-\partial^\mu T_{\mu\nu}^{(\textit{Max})}\ .
	\ee
In the massive fractonic theory we have seen that the CS-like term plays the role of ``intrinsic matter'' \eqref{Sm}, and is thus analogous to $S_J$ in \eqref{Stot}, in the same  way as $S_{fr}$ \eqref{fractterm} can be related to the invariant Maxwell action $S_{\textit{Max}}$. We therefore expect the appearance of an ``intrinsic force'' in analogy with \eqref{Fmax}. From the fractonic energy-moment tensor \eqref{T00fr}-\eqref{Tijfr},  using the Gauss \eqref{Gauss_vacuum}, Amp\`ere \eqref{Ampere_vacuum}, and Faraday \eqref{Faraday from Bianchi} laws, $i.e.$ going on-shell, we have
	\begin{align}
	\partial^\alpha T^{(\textit{fr})}_{\alpha 0} &=\frac{1}{2}E_{ij}J^{ij}_{\textsc{cs}}=0\label{dT0=0}\\
	\partial^\alpha T^{(\textit{fr})}_{\alpha i} &= F_i - f_i^{\textsc{cs}}\ ,\label{dTi}
\end{align}	
where \eqref{dT0=0} can be directly verified using \eqref{J}, and
\begin{align}
F_i &\equiv\frac{3}{4}B_i \partial^j B_j+\frac{1}{2} B_j\partial^j B_i  - \frac{1}{2}E^{jk}\partial_j E_{ik}\label{Fi}\\
f_i^{\textsc{cs}} &\equiv -\rho_{\textsc{cs}}^jE_{ij}+\frac{1}{2}\epsilon_{0ij}J^{jk}_{\textsc{cs}}B_k\label{lorentz force}\ .
\end{align}
In analogy with \eqref{Fmax}, \eqref{dT0=0} leads to
	\be\label{JE=0}
	f_0^{\textsc{cs}}=-\frac{1}{2}E_{ij}J^{ij}_{\textsc{cs}}=0 \ ,
	\ee
which means that the generalized current $J^{ij}_{\textsc{cs}}(x)$ \eqref{J} and the electric tensor field $E^{ij}(x)$ \eqref{def_E} are orthogonal to each other, which is typical of a Hall-like current \cite{Bertolini:2024yur,Prem:2017kxc}, and the fractonic power $f_0^{\textsc{cs}}(x)$ due to the CS-like term vanishes \eqref{JE=0}. Notice that the fact that the power $f_0^{\textsc{cs}}(x)$ \eqref{JE=0} vanishes is in agreement with the interpretation of the CS-like contribution as intrinsic matter. Indeed, since the power is referred to some kind of dispersion/energy transfer, we expect it to vanish in a process which is exclusively internal. On the other hand, we see that with respect to the massless case \eqref{breaking}, the divergence of the energy-moment tensor \eqref{dTi} has two contributions. The first $F_i(x)$ \eqref{Fi} is a consequence of the breaking of diffeomorphism invariance, and the second $f_i^{\textsc{cs}}(x)$ \eqref{lorentz force} is due to the presence of the CS-like term and can be recognized as a kind of  Lorentz-like force related to the (intrinsic) matter contribution. Interestingly, this force can also be rewritten in terms of the electric and magnetic fields as
	\be
	f_i^{\textsc{cs}}=-\frac1 4 m\left(2E_{ia}B^a+EB_i\right)\ ,
	\ee
with $E(x)\equiv \eta^{ij}E_{ij}(x)$, which makes explicit its gauge invariance and reminds the Lorentz force contribution (Eq.(28) of \cite{Brevik:2022gkt} for instance) of axion electrodynamics \cite{Sikivie:1983ip,Wilczek:1987mv,Sekine:2020ixs}.	

\section{Coupling to matter}

In presence of external matter the action can be written as
	\begin{equation}\label{Stot-frac}
	S_{tot} \equiv S_{inv} + S_J \ ,
	\end{equation}
where $S_{inv}$ is given by \eqref{Sinv},
	\begin{equation}
	S_J \equiv -\int \dd^3 x\, J^{\mu\nu}h_{\mu\nu}  \label{Sj}
	\end{equation}
and $J^{\mu\nu}(x)=J^{\nu\mu}(x)$ is the matter current. The on-shell EoM
	\begin{equation}
	\frac{\delta S_{tot}}{\delta h_{\alpha\beta}}=0
	\end{equation}
generalizes \eqref{eomSinv} to 
	\begin{align}
	- \partial_\mu F^{\alpha\beta\mu} = J^{\alpha\beta}_{\textsc{cs}}+ J^{\alpha\beta} \ ,\label{EoM_matter}
	\end{align}
where $J^{\mu\nu}_{\textsc{cs}}(x)$ is given by \eqref{def_J_cs}. Therefore, according to the discussion of the previous Sections, we can see that the MCS-like theory \eqref{Stot-frac} is characterized by two types of matter: an intrinsic one $J^{\mu\nu}_{\textsc{cs}}(x)$ \eqref{def_J_cs} and an external one $J^{\mu\nu}(x)$ \eqref{Sj}. Using \eqref{particular_solution}, the $00$-component of the on-shell EoM \eqref{EoM_matter}, implies
\begin{equation}
    J^{00}=0 \label{J00=0}\ .
\end{equation}
Moreover, from the EoM \eqref{EoM_matter}, one has
\begin{equation}
    \partial_\alpha \partial_\beta J^{\alpha\beta} =0\ ,
\end{equation}
which is the typical sign of a fractonic behaviour. In fact, as a consequence of \eqref{J00=0}, this nonstandard conservation equation can be rewritten in the same way as \eqref{cont-dipole}
	\begin{equation}
	\partial_0 \rho + \partial_i \partial_j J^{ij}=0 \ ,\label{fracton_continuity_eq}
	\end{equation}
where we defined the charge density
	\begin{equation}
	\rho \equiv 2\partial_i J^{0i} \label{def_rho_fractonic} \ ,
	\end{equation}
as in \eqref{rho_cs}. The fractonic behaviour emerges from  \eqref{fracton_continuity_eq}, since it implies the conservations of both the total charge
\be
	\partial_0 \int  d \Sigma\, \rho = -\int d \Sigma\, \partial_i \partial_j J^{ij}=0 \label{charge cons}
	\ee
and, up to boundary terms, the total dipole moment $D^k(t)$:
	\be\label{dipole cons}
	\partial_0 D^k \equiv \partial_0 \int d\Sigma\, d^k=   \partial_0 \int  d \Sigma\, x^k \rho = -\int d \Sigma\,x^k \partial_i \partial_j J^{ij}=0\ ,  
	\ee
where 
	\begin{equation}
	d^i \equiv -2 J^{i0}
	\end{equation}
is the dipole density. Notice however that, differently from the intrinsic case \eqref{cont-dipole}, which implies the conservation of ${D_{\textsc{cs}}}^i_{\ i}(t)$ \eqref{d0TrQ=0}, the continuity equation \eqref{fracton_continuity_eq} for the external matter does not imply the conservations of the trace of the quadrupole moment 
	\be
	D \equiv \int d\Sigma\, x^2 \rho\ ,
	\ee
since in this case $\mixt{J}{\mu}{\mu}(x)\neq 0$. Therefore the external matter contribution describes quasiparticles with different mobility: fractonic immobile charges and free dipoles, related to the conservation of charge \eqref{charge cons} and total dipole moment \eqref{dipole cons} only \cite{Pretko:2016kxt,Pretko:2016lgv}. Taking into account the definitions of the electric and magnetic-like fields $E^{ij}(x)$ \eqref{def_E} and $B^i(x)$ \eqref{def_B}, the nontrivial components of the EoM \eqref{EoM_matter} read
	\begin{itemize}
	\item $\alpha=0$, $\beta=i$
		\be
		\partial_j E^{ij}=-\rho^i_{\textsc{cs}} -d^i \label{EoM00_matter}\ ,
		\ee
	from which it follows that the Gauss constraint is
		\be
		\partial_i\partial_j E^{ij} = \rho_{\textsc{cs}} + \rho \label{Gauss_matter}\ ,
		\ee
where we used the definitions of intrinsic fractonic charge $\rho_{\textsc{cs}}(x)$ \eqref{Gauss_vacuum} and dipole $\rho^i_{\textsc{cs}}(x)$ \eqref{Gauss-vec} densities. Remembering that $\rho_{\textsc{cs}}(x)\propto \partial_i B^i(x)$ through \eqref{Gauss_vacuum}, the above equation implies that the fracton density $\rho(x)$ is a source for both generalized electric and magnetic fields, similarly to \cite{Pretko:2017xar}.
Moreover, from \eqref{EoM00_matter}, we observe that
		\begin{equation}
		-\int \dd \Sigma\, B^i = \frac{D^i}{m} \label{int_B_D_attachment}\ ,
		\end{equation}
is a higher-rank generalization of the relation which characterizes the ordinary abelian MCS theory \cite{Deser:1981wh,Deser:1982vy}
		\begin{equation}
		-\int \dd\Sigma\, B = \frac{Q}{m}\ ,
		\end{equation}
where $B(x)$ and $Q(t)$ are the ordinary magnetic field and the total electric charge  respectively. Notice that \eqref{int_B_D_attachment} can be seen an integrated version of the dipole-flux attachment relation that is found in the higher-rank CS theory for fractons \cite{Bertolini:2024yur}. As a consequence of the introduction of external matter, we also observe that from \eqref{Gauss_matter} and \eqref{int_xdotB=0} it follows that
		\begin{equation}
		\int \dd\Sigma\, x^2 \rho = 2\int \dd\Sigma\, E \ ,\label{trace_quadrupole}
		\end{equation}
which means that in this case, differently from the (vacuum) intrinsic CS-like matter case \eqref{trace_E_cs}, the trace of the electric field $E^{ij}(x)$ does not vanish globally. In other words, in presence of an external matter coupling, a longitudinal motion of the dipole $d^i(x)$ is allowed and associated to a change in the trace $E(x)$  \cite{Pretko:2017kvd}.
	\item $\alpha=i$, $\beta=j$
		\be
		\partial_0 E^{ij}-\frac{1}{2}\left(\epsilon^{0ki}\partial_k B^j + \epsilon^{0kj}\partial_k B^i\right)+\frac{1}{2}m \left(\epsilon^{0ik}\mixt{E}{j}{k}+\epsilon^{0jk}\mixt{E}{i}{k}\right)+J^{ij}=0\label{EoMij_matter}\ ,
		\ee
	which, by recalling the definition \eqref{J}, can also be written as
		\be
		-\partial_0 E^{ij}+\frac{1}{2}\left(\epsilon^{0ki}\partial_k B^j + \epsilon^{0kj}\partial_k B^i\right)=J^{ij}_{\textsc{cs}}+J^{ij}\label{EoMij_matter}\ .
		\ee
	\end{itemize}
When taking into account the external matter contribution \eqref{Sj}, the changes in the on-shell conservation of the total energy-moment tensor allow us to identify additional contributions to the power and force in \eqref{dT0=0} and \eqref{dTi}:
	\begin{align}
	\partial^\alpha T^{(\textit{fr})}_{\alpha 0} &=-f_0^{(\textsc{l})}\label{dT0=f0}\\
	\partial^\alpha T^{(\textit{fr})}_{\alpha i} &= F_i - f_i^{\textsc{cs}}- f_i^{(\textsc{l})}\ ,\label{dTi_matter}
	\end{align}	
where $F_i(x)$ is the diffeomorphism breaking term \eqref{Fi}, $f_i^{\textsc{cs}}(x)$ is the intrinsic Lorentz force \eqref{lorentz force} and
	\begin{align}
	f_0^{(\textsc{l})}&\equiv -\frac{1}{2}E_{ij}J^{ij}\label{power-matt}\\
	f_i^{(\textsc{l})} &\equiv -d^jE_{ij}+\frac1 2 \epsilon_{0ij}J^{jk}B_{k}
	\end{align}
are respectively the fractonic power and Lorentz-like force associated to the external fractonic matter. Notice that, due to the external matter contribution, the power $f_0^{(\textsc{l})}(x)$ \eqref{power-matt} does not vanish, differently from the contribution \eqref{JE=0} of the CS-like term.

\section{Conclusions}

The study of fractons has revealed novel and unconventional features in both high-energy and condensed matter physics, including restricted mobility, subdimensional symmetries, and emergent gauge structures. While much of the literature has focused on gapless or fractonic systems with limited mobility, in this article we have shown that the introduction of a finite mass offers a richer dynamical behaviour, allowing for the propagation of gapped modes while retaining key fractonic constraints. A consistent description of massive fractons is essential for extending the fracton paradigm to more general settings, including finite-temperature dynamics, real-time evolution, and potential experimental realizations. From a theoretical perspective, the formulation of a covariant theory of massive fractons with controlled gauge symmetry and dynamical DoF is a necessary step toward embedding fractonic models into relativistic and possibly gravitational frameworks.
In this work, we have introduced and explored a covariant, gauge-invariant theory for massive fractons in 3D, based on a symmetric rank-2 tensor gauge field. The cornerstone of the model is the presence of a CS-like term that is specific to three-dimensional spacetime and is constructed using the Levi-Civita tensor and the tensor field $h_{\mu\nu}(x)$. This term exhibits two key properties that form the core of our analysis and represent the main novelty of the paper.  
First, the CS-like term provides a mechanism of topological mass generation for the tensor gauge field $h_{\mu\nu}(x)$, similar in spirit to what happens in the MCS theory of Deser, Jackiw, and Templeton. However, unlike the MCS case -- which involves vector fields -- the mechanism here operates on a rank-2 symmetric tensor field, and the resulting mass term remains consistent with the underlying gauge symmetry, defined by longitudinal diffeomorphisms. In this sense, the analogy with the MCS mechanism is not just structural but it is also a conceptual extension to higher-rank fields, in a fully covariant way.
Crucially, the theory exhibits a well-defined and smooth massless limit. In fact, as the mass parameter $m \to 0$, the theory reduces to a covariant fracton model of the kind described in \cite{Bertolini:2022ijb}, where the dynamics is controlled solely by the kinetic (Maxwell-like) term. The massless limit preserves the gauge structure and the number of DoF, indicating that the introduction of the CS-like term enriches the physical content of the model without introducing pathologies or discontinuities, such as the vDVZ-type discontinuities observed in some massive spin-2 theories.
Second, and more remarkably, the same CS-like term plays a distinct and independent role: it acts as a source of intrinsic matter. Even in the absence of any external matter coupling, the Maxwell-like EoM acquire nontrivial right-hand sides in the form of current and charge densities that are completely determined by the presence of the CS-like term itself. These effective sources -- encoded in a conserved current $J_{\mu\nu}^{\text{cs}}(x)$ -- can be associated to fractonic behaviours, including Gauss and Amp\`ere-like equations, mobility constraints coming from the conservation of the dipole and of the trace of the quadrupole moment. This dual role of the CS-like term -- topological mass generator and internal source of fractonic matter -- is, to the best of our knowledge, unprecedented in the literature.
This intrinsic matter is not an auxiliary construction; it manifests itself in a physically meaningful way. In particular we have shown that one of the two dynamical DoF of the theory corresponds to a massive fractonic charge density $\rho_{\text{cs}}(x)$, which satisfies a Klein-Gordon equation. This identification provides a precise physical meaning to the massive excitation, which is thus not a generic tensor mode, but a propagating fracton, whose dynamics is governed by the internal structure of the theory itself. The second degree of freedom remains massless and is associated with the longitudinal sector of the spatial components $h_{ab}(x)$, consistently with the results found in the massless theory.
The DoF analysis confirms that the number of propagating modes remains unchanged, both in the massless and in the massive regime. This is a crucial consistency check. It supports the interpretation of the CS-like term as a quasi-topological contribution: although it is metric-dependent and hence not fully topological in the strict sense, it retains the key property of a genuine topological term, $i.e.$ it modifies the nature of the dynamics (introducing a mass) without increasing the number of physical DoF. This feature mirrors that of the MCS theory, where the CS term modifies the dispersion relation while preserving the single degree of freedom of the 3D photon.
Another important feature of our analysis is the identification of conservation laws -- such as those of the dipole and of the trace of the quadrupole moment -- associated with the intrinsic matter sector, which confirm the fractonic nature of the theory. These conservation laws are in fact responsible for the restricted mobility of the effective charges and currents. Remarkably, they emerge from the internal structure of the action, without the need for external source couplings. In particular, the tracelessness of the intrinsic current $J^{ij}_{\text{cs}}(x)$ ensures the conservation of the trace of the quadrupole moment, while the vanishing of the associated power $f_0^{\textsc{cs}}(x)$ implies that the intrinsic fractonic excitations behave as internal DoF with no net energy exchange with the environment. These properties further reinforce the interpretation of the CS-like term as an effective internal matter sector.
When external matter is introduced, the picture becomes even richer. The EoM acquire additional source terms, and the total energy-moment tensor receives contributions both from the intrinsic (CS-like) and extrinsic sectors. While the intrinsic sector continues to exhibit fully fractonic behaviours -- including restricted motion and conservation of both dipole and trace of the quadrupole moment -- the external sector allows for dipoles with complete mobility, and the quadrupole trace is no longer conserved. In this way, the theory describes a system with two coexisting fractonic sectors: one \textit{intrinsic}, constrained and massive; the other, \textit{external}, partially mobile and potentially gapless. This coexistence provides a versatile framework for modeling interacting fractonic systems and could serve as a starting point for more complex constructions, such as dualities with topological phases or gravitational analogues.
More precisely, the introduction of external matter enriches the intrinsic fractonic matter encoded in the CS-like term, with distinct physical features. The CS-like term describes the fractonic charge density $\rho_{\textsc{cs}}(x)$ as a massive propagating degree of freedom via equation \eqref{massiveKG}. Moreover, the intrinsic dipole-like current $J^{ij}_{\textsc{cs}}(x)$ 
 is such that the corresponding power $f_0^{\textsc{cs}}(x)$ vanishes,  and its tracelessness ensures the conservation of the trace of the quadrupole moment  $\partial_0 {D_{\textsc{cs}}}^i_{\ i}(t) = 0$. In contrast, the extrinsic dipole-like current $J^{ij}(x)$ contributes nontrivially to the power  $f_0^{(\textsc{l})}(x)$, and its trace enables longitudinal motion of the dipoles $d^i(x)$. Therefore, when external matter is present, the theory realizes two superimposed fractonic subsystems:
\begin{itemize}
\item an \textit{intrinsic} one with a massive, immobile fractonic charge $\rho_{\textsc{cs}}(x)$ and a lineon-like dipole $\rho_{\textsc{cs}}^i(x)$;
\item an \textit{external} one with immobile fractonic charge $\rho(x)$ and a dipole density $d^i(x)$ that can move both longitudinally and transversally.
\end{itemize}
In conclusion, the motivation for studying massive fractons lies not only in extending the fracton paradigm to more dynamical and realistic settings, but also in uncovering new mechanisms -- like the one presented here -- that unify mass generation and matter content within a single term. The model developed in this paper offers a controlled and robust arena in which these ideas can be investigated in depth, with possible applications ranging from effective theories of gapped fracton phases to covariant formulations of generalized symmetries.
Therefore, from a broader perspective, the theory developed here provides a field-theoretic framework for describing massive fractons in a fully covariant and gauge-invariant way. Fractons are typically studied in condensed matter contexts, often via lattice models or non-relativistic continuum theories. The present construction, in contrast, embeds fractonic behaviour into a relativistic field theory, thereby opening new possibilities for connecting fracton physics with high-energy and gravitational frameworks.

\appendix

\section{Computation of the propagators}\label{app-prop}

In moment space\footnote{The Fourier transform is defined $h_{\mu\nu}(x)=\int d^3p\  e^{ip_\lambda x^\lambda}\Tilde{h}_{\mu\nu}(p)$.}, the action $S$ \eqref{totactfract} reads:
	\begin{align}
	   S &= \int d^3p\left [
	   \tilde{h}_{\mu\nu}(p)
	   \left(
	   \eta^{\mu\alpha} p^\nu p^\beta  
	   -p^2\eta^{\mu\alpha}\eta^{\nu\beta}
	   -im p_\lambda \epsilon^{\mu\lambda\alpha} \eta^{\beta\nu}
	   \right)\tilde{h}_{\alpha\beta}(-p)\right.\nonumber\\
	   &\left.\qquad\qquad
	 + \tilde{b}(p)\left(-\kappa_0p^\alpha p^\beta- \kappa_1 p^2 \eta^{\alpha\beta}\right)\tilde{h}_{\alpha\beta}(-p)\right]\nonumber\\
	&= \int d^3p\ \tilde\phi_M(p)\tilde{K}^{MA}(p)\tilde\phi_A(-p)\ ,
	\label{momS}
	\end{align}
with
	\be
	{\tilde{K}}^{MA}(p) \equiv 
		\begin{bmatrix}
		\tilde{K}^{\mu\nu,\alpha\beta}(p) & \tilde{K}^{*\; \mu\nu}(p)\\ 
		\tilde{K}^{\alpha\beta}(p) & 0
		\end{bmatrix} \quad;\quad\tilde\phi_M\equiv\left(\tilde{h}_{\mu\nu}\ ,\ \tilde{b}\right)\ ,
	\label{defK}
	\ee
and 
	\begin{align}
	\tilde{K}^{\mu\nu,\alpha\beta}&= 
	\left[p^2\left(-A^{(0)}+\frac{1}{4}
	A^{(1)}
	\right){-\frac{m}{4}A^{(5)}}\right]^{\mu\nu,\alpha\beta}\\
	\tilde{K}^{\alpha\beta}&= -\frac{1}{2}\left(\kappa_0p^\alpha p^\beta+ \kappa_1 p^2 \eta^{\alpha\beta}\right)\ ,
	\label{Kexpansion}
	\end{align}
where we expanded the tensor $\tilde{K}^{\mu\nu,\alpha\beta}(p)$ on the $A^{(i)}$-basis \eqref{A0}-\eqref{A6} \cite{Amoretti:2013xya,Bertolini:2020hgr,Bertolini:2021iku}, and the fact that $h_{\mu\nu}(x)$ is a symmetric tensor field has been taken into account, for which the following symmetries hold
	\begin{align} 
	\tilde{K}^{\mu\nu,\alpha\beta}(p) &= \tilde{K}^{\nu\mu,\alpha\beta}(p) = \tilde{K}^{\mu\nu,\beta\alpha}(p)=\tilde{K}^{\alpha\beta,\mu\nu}(-p)=\tilde{K}^{*\ \alpha\beta,\mu\nu}(p) \\
	\tilde{K}^{\alpha\beta}(p) &= \tilde{K}^{\beta\alpha}(p)=\tilde{K}^{\alpha\beta}(-p) = \tilde{K}^{*\alpha\beta}(p) \ .
	\label{simmK}
	\end{align}
The matrix of the propagators in moment space is
	\be
	\tilde{\Delta}_{AP}(p) \equiv 
		\begin{bmatrix}
		 \tilde{\Delta}_{\alpha\beta,\rho\sigma}(p) & \tilde{\Delta}^*_{\alpha\beta}(p)\\ 
		  \tilde{\Delta}_{\rho\sigma}(p) & \tilde{\Delta}(p)
		\end{bmatrix} \ ,
	\label{defDelta}
	\ee
with
	\begin{align} 
	 \tilde{\Delta}_{\alpha\beta,\rho\sigma}(p)&\equiv \langle\Tilde{h}_{\alpha\beta}(p)\Tilde{h}_{\rho\sigma}(-p)\rangle  \\ 
	 \tilde{\Delta}_{\alpha\beta}(p)&\equiv \langle\Tilde{h}_{\alpha\beta}(p) \tilde{b}(-p)\rangle\\
	 \tilde{\Delta}(p)&\equiv \langle\tilde{b}(p) \tilde{b}(-p)\rangle\ ,
	\end{align}
and is defined such that
	\be
	{\tilde{K}}^{MA}\tilde{\Delta}_{AP}=
		\begin{bmatrix}
		 \mathcal{I}^{\mu\nu}_{\rho\sigma}& 0\\ 
		  0 & 1
		\end{bmatrix} \ ,
	\label{defprop}
	\ee
with
	\be
	\mathcal{I}^{\mu\nu}_{\rho\sigma}\equiv \frac{1}{2}\left(
	\delta^\mu_\rho \delta^\nu_\sigma + \delta^\mu_\sigma \delta^\nu_\rho
	\right)\ .
	\ee
On the basis $\{A^{(i)}\}$ \eqref{A0}-\eqref{A6} the propagator $\tilde{\Delta}_{\alpha\beta,\rho\sigma}(p)$ can be expanded as
	\be
	\tilde{\Delta}_{\alpha\beta,\rho\sigma}(p) = \sum_{i=0}^6 c_i(p) A^{(i)}_{\alpha\beta,\rho\sigma}(p)\ , 
	\label{Delta1base}
	\ee
and
	\be
	\tilde{\Delta}_{\alpha\beta}(p) = a_1(p)\eta_{\alpha\beta} + a_2(p)\frac{p_\alpha p_\beta}{p^2}\label{Delta2base} \ ,
	\ee
where $c_i(p)$ and $a_1(p)$, $a_2(p)$ are real functions. Our aim is to evaluate $c_i(p)$, $a_1(p)$, $a_2(p)$ from \eqref{defprop}, which explicitly reads
	\begin{align}\label{propagatorrelations}
	&\tilde{K}^{\mu\nu,\alpha\beta}\tilde{\Delta}_{\alpha\beta,\rho\sigma} + \tilde{K}^{*\; \mu\nu}\tilde{\Delta}_{\rho\sigma} =\mathcal{I}^{\mu\nu}_{\rho\sigma}\\
	&\tilde{K}^{\mu\nu,\alpha\beta}\tilde{\Delta}^*_{\alpha\beta} + \tilde{K}^{\mu\nu}\tilde{\Delta} = 0\label{propagatorrelation2}\\
	&\tilde{K}^{\alpha\beta}\tilde{\Delta}_{\alpha\beta,\rho\sigma} = 0 \label{propagatorrelation3}\\
	&\tilde{K}^{\alpha\beta}\tilde{\Delta}^*_{\alpha\beta} = 1 \ .\label{propagatorrelation4}
	\end{align}
We thus get
	\begin{align}
	-c_0+4mc_5&=\tfrac{1}{p^2}\quad \mbox{from } \eqref{propagatorrelations}
	\label{prop-eq1}\\
	\tfrac{1}{2}c_0-c_1-3mc_5+mc_6&=0\\
	-c_2+2mc_5-\tfrac{1}{2}\kappa_1a_2&=0\\
	c_3+2mc_5-\tfrac{1}{2}\kappa_0a_1&=0\\
	c_3+2mc_5+\tfrac{1}{2}\kappa_1a_1&=0\\
	2c_1+c_2-2mc_6-\tfrac{1}{2}\kappa_0a_2&=0\\
	p^2c_5+\tfrac{m}{4}c_0&=0\\
	\tfrac{1}{2}p^2c_5-\tfrac{1}{2}mc_1-\tfrac{1}{2}p^2c_6&=0\\[15px]
	\kappa_0c_0+4c_1(\kappa_0+\kappa_1)+c_2(\kappa_0+3\kappa_1)+c_4(\kappa_0+\kappa_1)&=0\quad\mbox{from }\eqref{propagatorrelation2}\\
	\kappa_1 c_0+(\kappa_0+\kappa_1)c_2+c_3(\kappa_0+3\kappa_1)&=0\\[15px]
	a_1+\tfrac{\kappa_1}{2}\tilde\Delta&=0\quad\mbox{from }\eqref{propagatorrelation3}
	\\
a_1-\tfrac{\kappa_0}{2}\tilde\Delta&=0\\[15px]
	-\tfrac{1}{2}[(\kappa_0+\kappa_1)a_2+(\kappa_0+3\kappa_1)a_1]&=\tfrac{1}{p^2}\quad\mbox{from }\eqref{propagatorrelation4}\label{prop-eq4}
		\end{align}	
The solutions to \eqref{defprop} are given by  
\begin{alignat}{3}
c_0&= -\frac{1}{p^2+m^2}&&\qquad c_1=- \frac{1}{2(p^2+m^2)} \\ 
c_2&=\frac{1}{2}\left[\frac{\kappa_0+3\kappa_1}{(\kappa_0+\kappa_1)p^2}-\frac{1}{p^2+m^2}\right]&&\qquad c_3=\frac{1}{2}\left( \frac{1}{p^2+m^2}-\frac{1}{p^2}\right)\\
c_4&=\frac{1}{2}\left[\frac{7}{p^2+m^2}-\frac{(\kappa_0+3\kappa_1)^2}{p^2(\kappa_0+\kappa_1)^2}\right]&&\qquad c_5=\frac{m}{4p^2(p^2+m^2)}\\
c_6&=\frac{m}{4p^2}\frac{3}{p^2+m^2}\ ,&&\\[10px]
a_1&=0\quad ;\quad a_2=-\frac{2}{p^2(\kappa_0 + \kappa_1)}\ ,&&\label{lastexgen}
\end{alignat}
and
	\be
	\tilde{\Delta}(p)=0\ .\label{propbb}
	\ee
The massless limit of the above solution gives the following nontrivial coefficients
	\begin{align}
	c_0&= -\frac{1}{p^2}\quad;\quad  c_1 =- \frac{1}{2p^2}\quad;\quad c_2= \frac{1}{p^2}\frac{\kappa_1}{\kappa_0+\kappa_1}\quad;\quad c_4 = \frac{1}{2p^2}\left[7-\frac{(\kappa_0+3\kappa_1)^2}{(\kappa_0+\kappa_1)^2}\right]\label{masslessprim}\\ 
	a_2&=-\frac{2}{p^2(\kappa_0 + \kappa_1)}\label{masslessult}\ .
	\end{align}
Explicitly, the nonvanishing propagators are
	\begin{align} 
	 \tilde{\Delta}_{\alpha\beta,\rho\sigma}(p)&
	 = \langle\Tilde{h}_{\alpha\beta}(p)\;\Tilde{h}_{\rho\sigma}(-p)\rangle\label{propprim}\\
	 &=\frac{1}{2(p^2 + m^2)}
	 \left[ -2A^{(0)}-A^{(1)}-A^{(2)}+A^{(3)}+7A^{(4)}+\frac{m}{2p^2}(A^{(5)}+3A^{(6)})\right]+\nonumber \\
	 &\quad+\frac{1}{2p^2}\left[\frac{(\kappa_0+3\kappa_1)}{(\kappa_0+\kappa_1)}A^{(2)}-A^{(3)}-\frac{(\kappa_0+3\kappa_1)^2}{(\kappa_0+\kappa_1)^2}A^{(4)}\right]_{\alpha\beta,\rho\sigma}\ ,\label{propult}
	\end{align}
where in \eqref{propprim} we isolated the poles contributions. The massive pole is at 
	\be
	 p^2 = -m^2\ .
	 \label{massivepole}
	\ee
	


\begin{thebibliography}{15}

\bibitem{Nandkishore:2018sel}
R.~M.~Nandkishore and M.~Hermele,
Ann. Rev. Condensed Matter Phys. \textbf{10} (2019), 295-313
doi:10.1146/annurev-conmatphys-031218-013604.

\bibitem{Pretko:2020cko}
M.~Pretko, X.~Chen and Y.~You,
Int. J. Mod. Phys. A \textbf{35} (2020) no.06, 2030003
doi:10.1142/S0217751X20300033.

\bibitem{Gromov:2022cxa}
A.~Gromov and L.~Radzihovsky,
Rev. Mod. Phys. \textbf{96} (2024) no.1, 011001
doi:10.1103/RevModPhys.96.011001.

\bibitem{Pretko:2016kxt}
M.~Pretko,
Phys. Rev. B \textbf{95} (2017) no.11, 115139
doi:10.1103/PhysRevB.95.115139.

\bibitem{Pretko:2016lgv}
M.~Pretko,
Phys. Rev. B \textbf{96} (2017) no.3, 035119
doi:10.1103/PhysRevB.96.035119.

\bibitem{Gromov:2018nbv}
A.~Gromov,
Phys. Rev. X \textbf{9}, no.3, 031035 (2019)
doi:10.1103/PhysRevX.9.031035.

\bibitem{Chamon:2004}
C.~Chamon,
Phys.\ Rev.\ Lett.\  {\bf 94} (2005) 040402
doi.org/10.1103/PhysRevLett.94.04040.

\bibitem{Vijay:2016phm}
S.~Vijay, J.~Haah and L.~Fu,
Phys. Rev. B \textbf{94}, no.23, 235157 (2016)
doi:10.1103/PhysRevB.94.235157.

\bibitem{Vijay:2015mka}
S.~Vijay, J.~Haah and L.~Fu,
Phys. Rev. B \textbf{92}, no.23, 235136 (2015)
doi:10.1103/PhysRevB.92.235136.

\bibitem{Haah:2011}
J.~Haah,
Phys.\ Rev.\ A {\bf 83} (2011) 042330
doi.org/10.1103/PhysRevA.83.042330.

\bibitem{Ma:2017aog}
H.~Ma, E.~Lake, X.~Chen and M.~Hermele,
Phys. Rev. B \textbf{95} (2017) no.24, 245126
doi:10.1103/PhysRevB.95.245126.

\bibitem{Brown:2019hxw}
B.~J.~Brown and D.~J.~Williamson,
Phys. Rev. Res. \textbf{2} (2020) no.1, 013303
doi:10.1103/PhysRevResearch.2.013303.

\bibitem{Pretko:2017kvd}
M.~Pretko and L.~Radzihovsky,
Phys. Rev. Lett. \textbf{120}, no.19, 195301 (2018)
doi:10.1103/PhysRevLett.120.195301.

\bibitem{Gromov:2017vir}
A.~Gromov,
Phys. Rev. Lett. \textbf{122} (2019) no.7, 076403
doi:10.1103/PhysRevLett.122.076403.

\bibitem{Gromov:2019waa}
A.~Gromov and P.~Sur\'owka,
SciPost Phys. \textbf{8} (2020) no.4, 065
doi:10.21468/SciPostPhys.8.4.065.

\bibitem{Caddeo:2022ibe}
A.~Caddeo, C.~Hoyos and D.~Musso,
Phys. Rev. D \textbf{106} (2022) no.11, L111903
doi:10.1103/PhysRevD.106.L111903
[arXiv:2206.12877 [cond-mat.str-el]].

\bibitem{Pena-Benitez:2023aat}
F.~Pe{\~n}a-Ben{\'\i}tez and P.~Salgado-Rebolledo,
JHEP \textbf{04} (2024), 009
doi:10.1007/JHEP04(2024)009.

\bibitem{Hartong:2024hvs}
J.~Hartong, G.~Palumbo, S.~Pekar, A.~P{\'e}rez and S.~Prohazka,
SciPost Phys. \textbf{18} (2025) no.1, 022
doi:10.21468/SciPostPhys.18.1.022.

\bibitem{Shirley:2018vtc}
W.~Shirley, K.~Slagle and X.~Chen,
SciPost Phys. \textbf{6} (2019) no.4, 041
doi:10.21468/SciPostPhys.6.4.041.

\bibitem{Pretko:2017fbf}
M.~Pretko,
Phys. Rev. D \textbf{96}, no.2, 024051 (2017)
doi:10.1103/PhysRevD.96.024051.

\bibitem{Pretko:2017xar}
M.~Pretko,
Phys. Rev. B \textbf{96} (2017) no.12, 125151
doi:10.1103/PhysRevB.96.125151.

\bibitem{Gaiotto:2014kfa}
D.~Gaiotto, A.~Kapustin, N.~Seiberg and B.~Willett,
JHEP \textbf{02}, 172 (2015)
doi:10.1007/JHEP02(2015)172.

\bibitem{Seiberg:2020bhn}
N.~Seiberg and S.~H.~Shao,
SciPost Phys. \textbf{10}, no.2, 027 (2021)
doi:10.21468/SciPostPhys.10.2.027

\bibitem{McGreevy:2022oyu}
J.~McGreevy,
Ann. Rev. Condensed Matter Phys. \textbf{14} (2023), 57-82
doi:10.1146/annurev-conmatphys-040721-021029.

\bibitem{Cordova:2022ruw}
C.~Cordova, T.~T.~Dumitrescu, K.~Intriligator and S.~H.~Shao,
``Snowmass White Paper: Generalized Symmetries in Quantum Field Theory and Beyond,''
[arXiv:2205.09545 [hep-th]].

\bibitem{Gromov:2020yoc}
A.~Gromov, A.~Lucas and R.~M.~Nandkishore,
Phys. Rev. Res. \textbf{2} (2020) no.3, 033124
doi:10.1103/PhysRevResearch.2.033124.

\bibitem{Doshi:2020jso}
D.~Doshi and A.~Gromov,
Commun Phys 4, 44 (2021)
doi:10.1038/s42005-021-00540-4.

\bibitem{Grosvenor:2021rrt}
K.~T.~Grosvenor, C.~Hoyos, F.~Pe{\~n}a-Ben{\'\i}tez and P.~Sur{\'o}wka,
Phys. Rev. Res. \textbf{3} (2021) no.4, 043186
doi:10.1103/PhysRevResearch.3.043186.


\bibitem{Blasi:2022mbl}
A.~Blasi and N.~Maggiore,
Phys. Lett. B \textbf{833} (2022), 137304
doi:10.1016/j.physletb.2022.137304.

\bibitem{Bertolini:2022ijb}
E.~Bertolini and N.~Maggiore,
Phys. Rev. D \textbf{106} (2022) no.12, 125008
doi:10.1103/PhysRevD.106.125008.

\bibitem{Bertolini:2023juh}
E.~Bertolini, A.~Blasi, A.~Damonte and N.~Maggiore,
Symmetry \textbf{15} (2023) no.4, 945
doi:10.3390/sym15040945.

\bibitem{Bertolini:2023sqa}
E.~Bertolini, N.~Maggiore and G.~Palumbo,
Phys. Rev. D \textbf{108} (2023) no.2, 025009
doi:10.1103/PhysRevD.108.025009.

\bibitem{Afxonidis:2023pdq}
E.~Afxonidis, A.~Caddeo, C.~Hoyos and D.~Musso,
Phys. Rev. D \textbf{109} (2024) no.6, 065013
doi:10.1103/PhysRevD.109.065013.

\bibitem{Bertolini:2024yur}
E.~Bertolini, A.~Blasi, N.~Maggiore and D.~S.~Shaikh,
JHEP \textbf{10} (2024), 232
doi:10.1007/JHEP10(2024)232.

\bibitem{Rovere:2024nwc}
D.~Rovere,
Phys. Rev. D \textbf{110} (2024) no.8, 8
doi:10.1103/PhysRevD.110.085012.

\bibitem{Bertolini:2025jul}
E.~Bertolini, A.~Blasi, M.~Carrega, N.~Maggiore and D.~S.~Shaikh,
Phys. Rev. B \textbf{111} (2025) no.8, 085126
doi:10.1103/PhysRevB.111.085126.

\bibitem{Bertolini:2025qcy}
E.~Bertolini, A.~Blasi and N.~Maggiore,
Eur. Phys. J. C \textbf{85} (2025) no.1, 68
doi:10.1140/epjc/s10052-025-13821-x.

\bibitem{Bertolini:2024apg}
E.~Bertolini and H.~Kim,
Phys. Rev. D \textbf{111} (2025) no.2, 025006
doi:10.1103/PhysRevD.111.025006.

\bibitem{Hinterbichler:2025ost}
K.~Hinterbichler and A.~Joyce,
``A Partially Massless Superconductor,''
[arXiv:2507.15932 [hep-th]].

\bibitem{Bulmash:2018lid}
D.~Bulmash and M.~Barkeshli,
Phys. Rev. B \textbf{97} (2018) no.23, 235112
doi:10.1103/PhysRevB.97.235112.

\bibitem{Ma:2018nhd}
H.~Ma, M.~Hermele and X.~Chen,
Phys. Rev. B \textbf{98} (2018) no.3, 035111
doi:10.1103/PhysRevB.98.035111.

\bibitem{Slagle:2017wrc}
K.~Slagle and Y.~B.~Kim,
Phys. Rev. B \textbf{96} (2017) no.19, 195139
doi:10.1103/PhysRevB.96.195139.

\bibitem{Ma:2020svo}
X.~Ma, W.~Shirley, M.~Cheng, M.~Levin, J.~McGreevy and X.~Chen,
Phys. Rev. B \textbf{105} (2022) no.19, 195124
doi:10.1103/PhysRevB.105.195124.

\bibitem{Chen:2023oov}
X.~Chen, H.~T.~Lam and X.~Ma,
 ``Ground State Degeneracy of Infinite-Component Chern-Simons-Maxwell Theories,''
 [arXiv:2306.00291 [cond-mat.str-el]].

\bibitem{Prem:2017kxc}
A.~Prem, M.~Pretko and R.~Nandkishore,
Phys. Rev. B \textbf{97} (2018) no.8, 085116
doi:10.1103/PhysRevB.97.085116.

\bibitem{Deser:1981wh}
S.~Deser, R.~Jackiw and S.~Templeton,
Annals Phys. \textbf{140} (1982), 372-411
[erratum: Annals Phys. \textbf{185} (1988), 406]
doi:10.1016/0003-4916(82)90164-6.

\bibitem{Deser:1982vy}
S.~Deser, R.~Jackiw and S.~Templeton,
Phys. Rev. Lett. \textbf{48}, 975-978 (1982)
doi:10.1103/PhysRevLett.48.975.

\bibitem{vanDam:1970vg}
H.~van Dam and M.~J.~G.~Veltman,
Nucl. Phys. B \textbf{22}, 397-411 (1970)
doi:10.1016/0550-3213(70)90416-5.

\bibitem{Zakharov:1970cc}
V.~I.~Zakharov,
JETP Lett. \textbf{12}, 312 (1970).

\bibitem{Blasi:2017pkk}
A.~Blasi and N.~Maggiore,
Eur. Phys. J. C \textbf{77}, no.9, 614 (2017)
doi:10.1140/epjc/s10052-017-5205-y
[arXiv:1706.08140 [hep-th]].

\bibitem{Blasi:2015lrg}
A.~Blasi and N.~Maggiore,
Class. Quant. Grav. \textbf{34}, no.1, 015005 (2017)
doi:10.1088/1361-6382/34/1/015005
[arXiv:1512.01025 [hep-th]].

\bibitem{Gambuti:2021meo}
G.~Gambuti and N.~Maggiore,
Eur. Phys. J. C \textbf{81}, no.2, 171 (2021)
doi:10.1140/epjc/s10052-021-08962-8
[arXiv:2102.10813 [gr-qc]].

\bibitem{Bertolini:2023wie}
E.~Bertolini and N.~Maggiore,
Phys. Rev. D \textbf{108}, no.10, 105012 (2023)
doi:10.1103/PhysRevD.108.105012
[arXiv:2310.20303 [hep-th]].

\bibitem{Bertolini:2025jov}
E.~Bertolini and G.~Palumbo,
Annals Phys. \textbf{480} (2025), 170138
doi:10.1016/j.aop.2025.170138.

\bibitem{Dalmazi:2020xou}
D.~Dalmazi and R.~R.~L.~d.~Santos,
Eur. Phys. J. C \textbf{81} (2021) no.6, 547
doi:10.1140/epjc/s10052-021-09297-0.

\bibitem{Bertolini:2025rhz}
E.~Bertolini, E.~Lui and N.~Maggiore,
Phys. Rev. D \textbf{112}, 044035 (2025)
doi:10.1103/db71-jwd2

\bibitem{Gambuti:2020onb}
G.~Gambuti and N.~Maggiore,
Phys. Lett. B \textbf{807}, 135530 (2020)
doi:10.1016/j.physletb.2020.135530
[arXiv:2006.04360 [gr-qc]].

\bibitem{Nakanishi:1966zz}
N.~Nakanishi,
Prog. Theor. Phys. \textbf{35} (1966), 1111-1116
doi:10.1143/PTP.35.1111.

\bibitem{Lautrup:1967zz} 
  B.~Lautrup,
  Kong.\ Dan.\ Vid.\ Sel.\ Mat.\ Fys.\ Med.\  {\bf 35}, no. 11 (1967).

\bibitem{Alvarez-Gaume:1989ldl}
L.~Alvarez-Gaume, J.~M.~F.~Labastida and A.~V.~Ramallo,
Nucl. Phys. B \textbf{334} (1990), 103-124
doi:10.1016/0550-3213(90)90658-Z.

\bibitem{Itzykson:1980rh}
C.~Itzykson and J.~B.~Zuber,
``Quantum Field Theory,''
McGraw-Hill, 1980,
ISBN 978-0-486-44568-7

\bibitem{Turner:2019wnh}
C.~Turner,
PoS {Modave2018}, 001 (2019)
doi:10.22323/1.349.0001.

\bibitem{Carroll:2004st}
S.~M.~Carroll,
``Spacetime and Geometry: An Introduction to General Relativity,''
Cambridge University Press, 2019,
ISBN 978-0-8053-8732-2, 978-1-108-48839-6, 978-1-108-77555-7
doi:10.1017/9781108770385

\bibitem{Weinberg:1995mt}
S.~Weinberg,
Cambridge University Press, 2005,
ISBN 978-0-521-67053-1, 978-0-511-25204-4
doi:10.1017/CBO9781139644167


\bibitem{Dunne:1998qy}
G.~V.~Dunne,
``Aspects of Chern-Simons theory,''
Contribution to: Les Houches Summer School in Theoretical Physics, Session 69: Topological Aspects of Low-dimensional Systems,
[arXiv:hep-th/9902115 [hep-th]].

\bibitem{Bertolini:2024jen}
E.~Bertolini and H.~Kim,
``Strings as Hyper-Fractons,''
[arXiv:2410.11678 [hep-th]].

\bibitem{Misner:1973prb}
C.~W.~Misner, K.~S.~Thorne and J.~A.~Wheeler,
``Gravitation,''
W. H. Freeman, 1973,
ISBN 978-0-7167-0344-0, 978-0-691-17779-3.

\bibitem{tong}
D.~Tong,
``Lectures on General Relativity,''
\url{http://www.damtp.cam.ac.uk/user/tong/gr.html}.

\bibitem{DEGROOT196877}
S.R.~de Groot and L.G.~Suttorp,
Physica {\bf 39}, 1 (1968), 77-83
doi:10.1016/0031-8914(68)90048-7.

\bibitem{Medina:2017mcd}
R.~Medina and J.~Stephany,
``The energy-moment tensor of electromagnetic fields in matter,''
[arXiv:1703.02109 [physics.class-ph]].

\bibitem{Brevik:2022gkt}
I.~H.~Brevik, I.~H.~Brevik, M.~M.~Chaichian and M.~M.~Chaichian,
Int. J. Mod. Phys. A \textbf{37} (2022) no.24, 2250151
[erratum: Int. J. Mod. Phys. A \textbf{37} (2022) no.36, 2250151]
doi:10.1142/S0217751X22501512.

\bibitem{Sikivie:1983ip}
P.~Sikivie,
``Experimental Tests of the Invisible Axion,''
Phys. Rev. Lett. \textbf{51} (1983), 1415-1417
[erratum: Phys. Rev. Lett. \textbf{52} (1984), 695]
doi:10.1103/PhysRevLett.51.1415.

\bibitem{Wilczek:1987mv}
F.~Wilczek,
Phys. Rev. Lett. \textbf{58} (1987), 1799
doi:10.1103/PhysRevLett.58.1799.

\bibitem{Rosenberg:2010ia}
G.~Rosenberg and M.~Franz,
``Witten effect in a crystalline topological insulator,''
Phys. Rev. B \textbf{82} (2010), 035105
doi:10.1103/PhysRevB.82.035105.

\bibitem{Chatzistavrakidis:2020wum}
A.~Chatzistavrakidis, G.~Karagiannis and P.~Schupp,
``Torsion-induced gravitational $\theta$ term and gravitoelectromagnetism,''
Eur. Phys. J. C \textbf{80} (2020) no.11, 1034
doi:10.1140/epjc/s10052-020-08600-9.

\bibitem{raab}
Raab, Roger E., and Owen L. de Lange, 
``Multipole Theory in Electromagnetism: Classical, quantum, and symmetry aspects, with applications'', 
International Series of Monographs on Physics (Oxford, 2004; online edn, Oxford Academic, 1 Sept. 2007), 
ISBN: 9780198567271
doi.org/10.1093/acprof:oso/9780198567271.001.0001

\bibitem{Sekine:2020ixs}
A.~Sekine and K.~Nomura,
J. Appl. Phys. \textbf{129} (2021) no.14, 141101
doi:10.1063/5.0038804.

\bibitem{Amoretti:2013xya}
A.~Amoretti, A.~Braggio, G.~Caruso, N.~Maggiore and N.~Magnoli,
Adv. High Energy Phys. \textbf{2014}, 635286 (2014)
doi:10.1155/2014/635286
[arXiv:1308.6674 [hep-th]].

\bibitem{Bertolini:2020hgr}
E.~Bertolini and N.~Maggiore,
Symmetry \textbf{12}, no.7, 1134 (2020)
doi:10.3390/sym12071134
[arXiv:2006.14902 [hep-th]].

\bibitem{Bertolini:2021iku}
E.~Bertolini, G.~Gambuti and N.~Maggiore,
Phys. Rev. D \textbf{104}, no.10, 105011 (2021)
doi:10.1103/PhysRevD.104.105011
[arXiv:2110.13203 [hep-th]].

\end{thebibliography}
\end{document}